\documentclass
[aps,pra,amsfonts,amssymb,twocolumn,amsmath,preprintnumbers,nofootinbib,floatfix,
showpacs]{revtex4-1}%
\usepackage[dvips]{graphics}
\usepackage{graphicx}
\usepackage{bm}
\usepackage{amsmath}
\usepackage{amsfonts}
\usepackage{amssymb}%
\providecommand{\U}[1]{\protect\rule{.1in}{.1in}}
\begin{document}

\title{The role of band-index-dependent transport relaxation times in anomalous Hall effect}
\author{Cong Xiao, Dingping Li, and Zhongshui Ma}
\affiliation{School of Physics, Peking University, Beijing 100871, China \\
Collaborative Innovation Center of Quantum Matter, Beijing, 100871, China}

\begin{abstract}
We revisit model calculations of the anomalous Hall effect (AHE) and show that, in isotropic Rashba-coupled two-dimensional electron gas (2DEG) with pointlike potential impurities, the full solution of the semiclassical Boltzmann equation (SBE) may differ from the widely-used $1/\tau^{||}$\ $\&$%
\ $1/\tau^{\perp}$ solution [Phys. Rev. B 68, 165311 (2003)]. Our approach to AHE is analogous to the SBE-based analysis of the anisotropic magnetoresistance leading to an integral equation for the distribution function
[Phys. Rev. B 79, 045427 (2009)] but in the present case, we reduce the description to band-index-dependent transport relaxation times. When both Rashba bands are partially occupied, these are
determined by solving a system of linear equations. Detailed calculations show that, for intrinsic and hybrid skew scatterings the difference between $1/\tau^{||}$\ $\&$\ $1/\tau^{\perp}$ and the full
solution of SBE is notable for large Fermi energies. For coordinate-shift effects, the side-jump velocity acquired in the inter-band elastic scattering process is shown to be more important for larger Rashba coupling and may even exceed the intra-band one for the outer Rashba band. The coordinate-shift contribution to AHE in the considered case notably differs from that in the limit of smooth disorder potential analyzed before.
\end{abstract}


\maketitle

\section{Introduction}

The extensive researches on the anomalous Hall effect (AHE) in the past twenty
years have promoted some important improvements in the physical understanding
and theoretical tools \cite{Nagaosa2010}. The importance of the inter-band
coherence effects in the topologically non-trivial band structures
\cite{Niu2010} has been highlighted theoretically
\cite{Sinitsyn2007,Kovalev2010}. Some simple but topologically non-trivial
band models, such as the two-dimensional (2D) massive Dirac model
\cite{SinitsynPRL2006,Sinitsyn2007,Yang2011} and 2D spin-polarized Rashba
model
\cite{Culcer2003,Dugaev2005,Liu,Inoue2006,Sinitsyn2005,Onoda,Kovalev2008,Kovalev2009,Kovalev2010,Nunner2007,Borunda2007,Nunner2008}%
, have been studied for analytical account of both the intrinsic and extrinsic
mechanisms for AHE. While the intrinsic mechanism stems solely from the
nontrivial band structure, the extrinsic mechanism relies on the existence of
disorder and is the sum of two contributions known as skew scattering and
side-jump. In the language of semiclassical Boltzmann equation (SBE) framework
\cite{Sinitsyn2008}, the side-jump contribution arises from the
coordinate-shift effects \cite{Sinitsyn2005,Sinitsyn2006}, whereas the skew
scattering originates from the scattering asymmetry when the scattering rate
is calculated beyond the lowest Born approximation. Other than the
conventional skew scattering due to the non-Gaussian third-order disorder
correlation, other two skew scattering contributions have been recently
proposed: the intrinsic (disorder-independent) skew scattering
\cite{Sinitsyn2007,Czaja2014,Kovalev2008,Kovalev2009} due to the Gaussian
disorder and hybrid skew scattering \cite{Kovalev2008,Kovalev2009} due to the
fourth-order non-Gaussian disorder correlation.

The SBE approach to AHE, developed by Sinitsyn et al. in a series of excellent
papers \cite{Sinitsyn2005,Sinitsyn2006,Sinitsyn2007,Sinitsyn2008,Nagaosa2010},
is physically more transparent than quantum-mechanical Kubo-Streda
\cite{Dugaev2005,Sinitsyn2007,Nunner2007,Ebert2010,Ebert2015,Yang2011,Hou2015}
and multiple-band Keldysh \cite{Liu,Onoda,Kovalev2008,Kovalev2009,Kovalev2010}
approaches. For the 2D massive Dirac model, the equivalence between the
Kubo-Streda and SBE approaches under the non-crossing approximation has been
shown \cite{Sinitsyn2007}. However, for the 2D spin-polarized Rashba model,
although the equivalence between Kubo-Streda and Keldysh formalisms has been
explicitly shown \cite{Kovalev2009}, existing SBE calculations
\cite{Sinitsyn2005,Borunda2007} do not produce fully same results for
side-jump and skew scattering as quantum-mechanical transport theories
\cite{Kovalev2008,Kovalev2009,Kovalev2010}. Sinitsyn et al.
\cite{Sinitsyn2005} analyzed the coordinate-shift effects in the limit of
smooth disorder potential, showing that the contribution of coordinate-shift
has the same sign as the intrinsic one for sufficiently large Fermi energy. In
the limit of smooth disorder potential, the intra-band small-angle scattering
dominates the electron-impurity elastic scattering processes, leaving little
room for the intra-band scattering with large momentum transfer and inter-band
elastic scattering. However, the latter two elastic scattering processes may
be important in the case of short-range disorder potential. Because recent
convincing Kubo-Streda \cite{Nunner2007,Inoue2006} and Keldysh calculations
\cite{Onoda,Kovalev2008,Kovalev2009,Kovalev2010} assume pointlike potential
impurities, the comparison of SBE and these calculations calls for a SBE study
in the presence of pointlike potential impurities. The work of Borunda et al.
\cite{Borunda2007} is an effort towards this direction, showing that when both
Rashba bands are partially occupied the conventional skew scattering vanishes
in both the SBE and Kubo-Streda calculations. However, the SBE calculation of
coordinate-shift, intrinsic skew scattering and hybrid skew scattering were
not carried out in this work. On the other hand, this work employed the
$1/\tau^{||}$\ $\&$\ $1/\tau^{\perp}$ solution
\cite{Schliemann2003,Vyborny2009} to the SBE for the distribution functions
responsible for the longitudinal and skew scattering transport. The validity
of $1/\tau^{||}$\ $\&$\ $1/\tau^{\perp}$ solution has been verified in the 2D
massive Dirac model \cite{Sinitsyn2007}. However, as will be shown in the
present paper, this solution is not valid in a spin-polarized Rashba 2DEG with
pointlike potential impurities when both Rashba bands are partially occupied.

In the present paper, we analyze the solution of SBE in isotropic 2DEG in the
presence of static impurities. When considering longitudinal and
skew-scattering-induced Hall transport, for multiple-Fermi-circle 2DEG we show
that the solution may be different from the $1/\tau^{||}$\ $\&$\ $1/\tau
^{\perp}$ solution if a decoupling condition is unsatisfied. We prove
that this decoupling condition is unsatisfied in a spin-polarized Rashba
2DEG with pointlike potential impurities when both Rashba bands are partially
occupied. Focusing on this particular case, detailed calculations show that,
for intrinsic and hybrid skew scatterings, the deviation of $1/\tau^{||}$%
\ $\&$\ $1/\tau^{\perp}$ solution from our solution is notable for large
Fermi energies. For coordinate-shift effects, the side-jump velocity
acquired in the inter-band elastic scattering process is shown to
be more important for larger Rashba spin-orbit coupling. Especially, the
inter-band side-jump velocity may exceed the intra-band one for the outer
Rashba band. The coordinate-shift contribution to the AHE in the case of
pointlike impurities is significantly different from that in the limit of
smooth disorder potential analyzed by Sinitsyn et al. \cite{Sinitsyn2005}.

The paper is organized as follows. To be self-contained, Sec. II briefly
sketches the basic framework of the modern SBE theory. Sec. III presents
general discussion on the solution of SBE in isotropic multiple-Fermi-circle
2DEG, whereas Sec. IV focuses on concrete calculations of AHE in a
spin-polarized Rashba 2DEG. Sec. V concludes the paper by discussing the
difference between our solution of the SBE and the $1/\tau^{||}$%
\ $\&$\ $1/\tau^{\perp}$ solution. The Appendix contains the main steps of
calculating scattering rates.

\section{Basic framework}

The basic framework of SBE approach in isotropic electron gases has been
formulated by Sinitsyn \cite{Sinitsyn2008}, here we sketch it very briefly.
The starting point is the SBE describing the evolution of semiclassical
distribution function (DF) of electron wavepackets in the presence of a weak
uniform electric-field $\mathbf{E}$ and diluted static impurities. For
isotropic band structures, the linearized steady-state SBE can be formulated
as \cite{Sinitsyn2007}%
\begin{equation}
\mathbf{F}\cdot\mathbf{v}_{l}^{0}\frac{\partial f^{0}}{\partial\epsilon_{l}%
}=-\sum_{l^{\prime}}\omega_{l,l^{\prime}}\left(  g_{l}-g_{l^{\prime}}\right)
, \label{SBE-n-origin}%
\end{equation}%
\begin{equation}
\mathbf{F}\cdot\mathbf{v}_{l}^{sj}\frac{\partial f^{0}}{\partial\epsilon_{l}%
}=\sum_{l^{\prime}}\omega_{l,l^{\prime}}^{2s}\left(  g_{l}^{adis}%
-g_{l^{\prime}}^{adis}\right)  . \label{SBE-a-origin}%
\end{equation}
Here $l=\left(  \eta,\mathbf{k}\right)  $ denotes the eigenstate with $\eta$
the band index and $\mathbf{k}$ the momentum, $\mathbf{F}=e\mathbf{E}$ is the
driving force, $\mathbf{v}_{l}^{0}=\partial\epsilon_{l}/\hbar\partial
\mathbf{k}$ is the group velocity, $\omega_{l,l^{\prime}}$ the transition rate
from state $l^{\prime}$ to $l$ due to elastic electron-impurity scattering.
The scattering rate is determined by the golden rule $\omega_{l,l^{\prime}%
}=\frac{2\pi}{\hbar}\left\langle \left\vert T_{l,l^{\prime}}\right\vert
^{2}\right\rangle \delta\left(  \epsilon_{l^{\prime}}-\epsilon_{l}\right)  $
where $\left\langle ..\right\rangle $ stands for the disorder configuration
average and $T_{l,l^{\prime}}$ the elements of T-matrix (details in appendix).
To capture the skew scattering, the T-matrix expansion beyond the 1st Born
order is needed and then the principle of microscopic detailed balance breaks
down $\omega_{l,l\prime}\neq\omega_{l\prime,l}$. One can then define the
symmetric and anti-symmetric parts of scattering rate as: $\omega_{l\prime
,l}^{s\left(  a\right)  }\equiv\frac{1}{2}\left(  \omega_{l\prime,l}\pm
\omega_{l,l\prime}\right)  $. For simple isotropic bands $\sum_{l\prime}%
\omega_{l\prime,l}^{a}=0$ leads to the form of collision term in Eqs.
(\ref{SBE-n-origin}) and (\ref{SBE-a-origin}) \cite{Nagaosa2010}. The
semiclassical DF is decomposed into $f_{l}=f_{l}^{0}+g_{l}+g_{l}^{adis}$
around the equilibrium Fermi-Dirac DF $f^{0}$\ with $g_{l}$ equilibrating the
electron wavepacket acceleration resulted by the driving force between
successive scattering events and the anomalous DF $g_{l}^{adis}$ describing
the effect of external fields working during the coordinate-shift process.
Usually the coordinate-shift effect is dealt with in the 1st Born
approximation \cite{Sinitsyn2008}, thus in Eq. (\ref{SBE-a-origin})
$\omega_{l,l\prime}$ is approximated by its first Born approximation
$\omega_{l\prime l}^{2s}$. $\mathbf{v}_{l}^{sj}\equiv\sum_{l\prime}%
\omega_{l\prime,l}^{2s}\delta\mathbf{r}_{l\prime,l}$ is the side-jump velocity
\cite{Sinitsyn2005,Sinitsyn2006}, where $\delta\mathbf{r}_{l^{\prime},l}%
$\ denotes the coordinate-shift \cite{Sinitsyn2006} in the scattering process
which scatters the electron in state $l$ into $l^{\prime}$ and reads
$\delta\mathbf{r}_{l\prime,l}=\langle u_{l\prime}|i\partial_{\mathbf{k}%
^{\prime}}|u_{l\prime}\rangle-\langle u_{l}|i\partial_{\mathbf{k}}%
|u_{l}\rangle-\mathbf{\hat{D}}\arg\langle u_{l\prime}|u_{l}\rangle$ for
spin-independent scalar disorder in the 1st Born approximation
\cite{Sinitsyn2006}. Here $|u_{l}\rangle$ is defined via $|l\rangle
=|\mathbf{k}\rangle|u_{l}\rangle$, denoting the eigenstate related to the
internal degrees of freedom, and $\mathbf{\hat{D}}=\partial_{\mathbf{k}%
^{\prime}}+\partial_{\mathbf{k}}$.

\section{General Analysis in isotropic 2DEG}

Equations (\ref{SBE-n-origin}) and (\ref{SBE-a-origin}) can be re-expressed as%
\begin{gather}
\mathbf{F}\cdot\mathbf{v}_{\eta}^{0}\left(  \epsilon,\phi\right)
\frac{\partial f^{0}}{\partial\epsilon}=-\sum_{\eta^{\prime}}\int\frac
{d\phi^{\prime}}{2\pi}\omega_{\eta\phi,\eta^{\prime}\phi^{\prime}}\left(
\epsilon\right) \nonumber\\
\times\left[  g_{\eta}\left(  \epsilon,\vartheta\left(  \mathbf{v}_{\eta}%
^{0}\left(  \epsilon,\phi\right)  \right)  \right)  -g_{\eta^{\prime}}\left(
\epsilon,\vartheta\left(  \mathbf{v}_{\eta^{\prime}}^{0}\left(  \epsilon
,\phi^{\prime}\right)  \right)  \right)  \right]  \label{SBE-n-origin-1}%
\end{gather}
and%
\begin{gather}
\mathbf{F}\cdot\mathbf{v}_{\eta}^{sj}\left(  \epsilon,\phi\right)
\frac{\partial f^{0}}{\partial\epsilon}=\sum_{\eta^{\prime}}\int\frac
{d\phi^{\prime}}{2\pi}\omega_{\eta\phi,\eta^{\prime}\phi^{\prime}}^{2s}\left(
\epsilon\right) \label{SBE-a-origin-1}\\
\times\left[  g_{\eta}^{adis}\left(  \epsilon,\vartheta\left(  \mathbf{v}%
_{\eta}^{sj}\left(  \epsilon,\phi\right)  \right)  \right)  -g_{\eta^{\prime}%
}^{adis}\left(  \epsilon,\vartheta\left(  \mathbf{v}_{\eta^{\prime}}%
^{sj}\left(  \epsilon,\phi^{\prime}\right)  \right)  \right)  \right]
,\nonumber
\end{gather}
respectively. Here the energy-integrated transition rate%
\begin{equation}
\omega_{\eta\phi,\eta^{\prime}\phi^{\prime}}\left(  \epsilon\right)
=N_{\eta^{\prime}}\left(  \epsilon\right)  \int d\epsilon_{l\prime}%
\omega_{l,l\prime},
\end{equation}
is introduced, $\epsilon=\epsilon_{l}$. $l=\left(  \epsilon,\eta,\phi\right)
$ is the eigenstate index, $\phi$ is the polar angle of 2D momentum, $N_{\eta
}\left(  \epsilon\right)  $ denotes the density of states (DOS),
$\vartheta\left(  \mathbf{v}_{\eta}^{0\left(  sj\right)  }\right)  $ denotes
the angle between $\mathbf{v}_{\eta}^{0\left(  sj\right)  }$ and the external
field. In isotropic bands (assuming electron-like dispersion curve)
$\mathbf{v}_{\eta}^{0}\left(  \epsilon,\phi\right)  \propto\mathbf{k}_{\eta
}\left(  \epsilon\right)  $. Whereas generally one can expect $\mathbf{v}%
_{\eta}^{sj}\left(  \epsilon,\phi\right)  \propto\mathbf{\hat{z}}%
\times\mathbf{k}_{\eta}\left(  \epsilon\right)  $ due to the chirality of
side-jump velocity.

In isotropic multi-band 2DEG with isotropic scatterers, the normal part
$g_{\eta}$ of out-of-equilibrium DF can be described by introducing the
isotropic longitudinal and transverse transport relaxation times for electrons
on iso-energy circles with energy $\epsilon$ in the $\eta$ band as%
\begin{gather}
g_{\eta}\left(  \epsilon,\vartheta\left(  \mathbf{k}_{\eta}\left(
\epsilon\right)  \right)  \right)  =\left(  -\partial_{\epsilon}f^{0}\right)
\nonumber\\
\times\left[  \mathbf{F}\cdot\mathbf{v}_{\eta}^{0}\left(  \epsilon
,\phi\right)  \tau_{\eta}^{L}\left(  \epsilon\right)  +\left(  \mathbf{\hat
{z}\times F}\right)  \cdot\mathbf{v}_{\eta}^{0}\left(  \epsilon,\phi\right)
\tau_{\eta}^{sk}\left(  \epsilon\right)  \right]  . \label{DF}%
\end{gather}
$\tau_{\eta}^{L}\left(  \epsilon\right)  $ describes the longitudinal
transport along the driving force; whereas $\tau_{\eta}^{sk}\left(
\epsilon\right)  $ is responsible for the skew-scattering-induced Hall
transport. Similarly, the anomalous DF is described by%
\begin{equation}
g_{\eta}^{adis}\left(  \epsilon,\vartheta\left(  \mathbf{v}_{\eta}^{sj}\left(
\epsilon,\phi\right)  \right)  \right)  =\partial_{\epsilon}f^{0}%
\mathbf{F}\cdot\mathbf{v}_{\eta}^{sj}\left(  \epsilon,\phi\right)  \tau_{\eta
}^{sj}\left(  \epsilon\right)  , \label{DF-adis}%
\end{equation}
where $\tau_{\eta}^{sj}$ has the dimension of time and may also be band-dependent.\

Because all elastic scattering events occur in the iso-energy circles with
energy $\epsilon$, in the following derivation we suppress the variable
$\epsilon$ in the arguments of DF, transport time, scattering rate, group
velocity and so on. Substituting Eq. (\ref{DF})\ into (\ref{SBE-n-origin-1})
yields two equations:%
\begin{align}
\frac{1}{\tau_{\eta}^{L}}  &  =\sum_{\eta^{\prime}}\int\frac{d\phi^{\prime}%
}{2\pi}\omega_{\eta\phi,\eta^{\prime}\phi^{\prime}}\left[  1-\cos\left(
\phi^{\prime}-\phi\right)  \frac{v_{\eta^{\prime}}^{0}}{v_{\eta}^{0}}%
\frac{\tau_{\eta^{\prime}}^{L}}{\tau_{\eta}^{L}}\right. \nonumber\\
&  \left.  -\sin\left(  \phi^{\prime}-\phi\right)  \frac{v_{\eta^{\prime}}%
^{0}}{v_{\eta}^{0}}\frac{\tau_{\eta^{\prime}}^{sk}}{\tau_{\eta}^{L}}\right]  ,
\label{SBE-1}%
\end{align}%
\begin{align}
0  &  =\sum_{\eta^{\prime}}\int\frac{d\phi^{\prime}}{2\pi}\omega_{\eta
\phi,\eta^{\prime}\phi^{\prime}}\left[  \sin\left(  \phi^{\prime}-\phi\right)
\frac{v_{\eta^{\prime}}^{0}}{v_{\eta}^{0}}\frac{\tau_{\eta^{\prime}}^{L}}%
{\tau_{\eta}^{L}}+\frac{\tau_{\eta}^{sk}}{\tau_{\eta}^{L}}\right. \nonumber\\
&  \left.  -\cos\left(  \phi^{\prime}-\phi\right)  \frac{v_{\eta^{\prime}}%
^{0}}{v_{\eta}^{0}}\frac{\tau_{\eta^{\prime}}^{sk}}{\tau_{\eta}^{L}}\right]  ,
\label{SBE-2}%
\end{align}
where $\left\vert \mathbf{v}_{\eta}^{0}\left(  \epsilon,\phi\right)
\right\vert \equiv v_{\eta}^{0}\left(  \epsilon\right)  $. Substituting Eq.
(\ref{DF-adis})\ into (\ref{SBE-a-origin-1}) yields%

\begin{equation}
\frac{1}{\tau_{\eta}^{sj}}=\sum_{\eta^{\prime}}\int\frac{d\phi^{\prime}}{2\pi
}\omega_{\eta\phi,\eta^{\prime}\phi^{\prime}}^{2s}\left[  1-\frac
{\mathbf{F}\cdot\mathbf{v}_{\eta^{\prime}}^{sj}\left(  \phi^{\prime}\right)
}{\mathbf{F}\cdot\mathbf{v}_{\eta}^{sj}\left(  \phi\right)  }\frac{\tau
_{\eta^{\prime}}^{sj}}{\tau_{\eta}^{sj}}\right]  . \label{SBE-a-Born-1}%
\end{equation}
The second term on the right-hand-side (rhs) is proportional to $\cos\left(
\phi^{\prime}-\phi\right)  v_{\eta^{\prime}}^{sj}/v_{\eta}^{sj}$, but the
side-jump velocities on different Fermi circles may have preferences for
different chirality (left or right).

\subsection{Case of single Fermi circle}

In the case of only single Fermi circle for a given Fermi energy, $\tau_{\eta
}^{L\left(  sk\right)  }$ and $\tau_{\eta^{\prime}}^{L\left(  sk\right)  }%
$\ are decoupled ($\eta^{\prime}\neq\eta$) in Eqs. (\ref{SBE-1}) and
(\ref{SBE-2}), which can thus be greatly simplified into
\begin{align}
\frac{1}{\tau_{\eta}^{L}}  &  =\int\frac{d\phi^{\prime}}{2\pi}\omega_{\eta
\phi,\eta\phi^{\prime}}\left\{  \left[  1-\cos\left(  \phi^{\prime}%
-\phi\right)  \right]  \right. \nonumber\\
&  \left.  -\sin\left(  \phi^{\prime}-\phi\right)  \frac{\tau_{\eta}^{sk}%
}{\tau_{\eta}^{L}}\right\}  , \label{SBE-1-S}%
\end{align}%
\begin{align}
0  &  =\int\frac{d\phi^{\prime}}{2\pi}\omega_{\eta\phi,\eta\phi^{\prime}%
}\left\{  \sin\left(  \phi^{\prime}-\phi\right)  \right. \nonumber\\
&  \left.  +\left[  1-\cos\left(  \phi^{\prime}-\phi\right)  \right]
\frac{\tau_{\eta}^{sk}}{\tau_{\eta}^{L}}\right\}  . \label{SBE-2-S}%
\end{align}
One finds that two relaxation-time like quantities defined on each isotropic
Fermi circle as \cite{SinitsynPRL2006,Sinitsyn2007}
\begin{equation}
\frac{1}{\tau_{\eta}^{\Vert}}=\int\frac{d\phi^{\prime}}{2\pi}\omega_{\eta
\phi,\eta\phi^{\prime}}\left[  1-\cos\left(  \phi^{\prime}-\phi\right)
\right]
\end{equation}
and
\begin{equation}
\frac{1}{\tau_{\eta}^{\bot}}=\int\frac{d\phi^{\prime}}{2\pi}\omega_{\eta
\phi,\eta\phi^{\prime}}\sin\left(  \phi-\phi^{\prime}\right)
\end{equation}
solve Eqs. (\ref{SBE-1-S}) and (\ref{SBE-2-S}):%
\begin{equation}
\tau_{\eta}^{L}=\frac{\tau_{\eta}^{||}}{1+\left(  \frac{\tau_{\eta}^{||}}%
{\tau_{\eta}^{\perp}}\right)  ^{2}},\text{ \ }\tau_{\eta}^{sk}=\frac
{\tau_{\eta}^{\perp}}{1+\left(  \frac{\tau_{\eta}^{\perp}}{\tau_{\eta}^{||}%
}\right)  ^{2}}. \label{complete}%
\end{equation}
This solution coincides with the \textquotedblleft$1/\tau^{||}$\ $\&$%
\ $1/\tau^{\perp}$ approach\textquotedblright\ proposed by Schliemann and Loss
\cite{Schliemann2003,Vyborny2009}. Thereby, the validity of \textquotedblleft%
$1/\tau^{||}$\ $\&$\ $1/\tau^{\perp}$ approach\textquotedblright\ is confirmed
in isotropic single-Fermi-circle 2DEG. The n-doped 2D massive Dirac system is
just a single-Fermi-circle 2DEG, and the application of $1/\tau^{||}$%
\ $\&$\ $1/\tau^{\perp}$ solution in this model is
successful\cite{SinitsynPRL2006,Sinitsyn2007}.

Meanwhile, $\tau_{\eta}^{sj}$ and $\tau_{\eta^{\prime}}^{sj}$\ are decoupled
for $\eta^{\prime}\neq\eta$ in Eq. (\ref{SBE-a-Born-1}), which thus reduces to%
\begin{equation}
\frac{1}{\tau_{\eta}^{sj}}=\int\frac{d\phi^{\prime}}{2\pi}\omega_{\eta
\phi,\eta^{\prime}\phi^{\prime}}^{2s}\left[  1-\cos\left(  \phi^{\prime}%
-\phi\right)  \right]  .
\end{equation}
We see that in the case of single Fermi circle, $\tau_{\eta}^{sj}$ is just the
longitudinal transport time $\tau_{\eta}^{L}$ calculated in the first Born
order. This relation can be easily verified in the 2D massive Dirac model
\cite{Sinitsyn2007}.

\subsection{Case of multiple Fermi circles}

In 2DEG with more than single Fermi circle for a given Fermi energy, the
possible presence of inter-band elastic scattering complicates the solution of
Boltzmann equations because in this case $\tau_{\eta}^{L\left(  sk\right)  }$
and $\tau_{\eta^{\prime}}^{L\left(  sk\right)  }$\ are coupled ($\eta^{\prime
}\neq\eta$) in Eqs. (\ref{SBE-1}) and (\ref{SBE-2}). Noticing that for
isotropic bands $\int\frac{d\phi^{\prime}}{2\pi}\omega_{\eta\phi,\eta^{\prime
}\phi^{\prime}}^{a}=0$, Eqs. (\ref{SBE-1}) and (\ref{SBE-2}) reduce to the
following two coupled equations%
\begin{align}
1  &  =\sum_{\eta^{\prime}}\int\frac{d\phi^{\prime}}{2\pi}\omega_{\eta
\phi,\eta^{\prime}\phi^{\prime}}^{s}\left[  \tau_{\eta}^{L}-\cos\left(
\phi^{\prime}-\phi\right)  \frac{v_{\eta^{\prime}}^{0}}{v_{\eta}^{0}}%
\tau_{\eta^{\prime}}^{L}\right] \nonumber\\
&  -\sum_{\eta^{\prime}}\int\frac{d\phi^{\prime}}{2\pi}\omega_{\eta\phi
,\eta^{\prime}\phi^{\prime}}^{a}\sin\left(  \phi^{\prime}-\phi\right)
\frac{v_{\eta^{\prime}}^{0}}{v_{\eta}^{0}}\tau_{\eta^{\prime}}^{sk},
\label{SBE-1-2}%
\end{align}
and%
\begin{align}
0  &  =\sum_{\eta^{\prime}}\int\frac{d\phi^{\prime}}{2\pi}\omega_{\eta
\phi,\eta^{\prime}\phi^{\prime}}^{s}\left[  \tau_{\eta}^{sk}-\cos\left(
\phi^{\prime}-\phi\right)  \frac{v_{\eta^{\prime}}^{0}}{v_{\eta}^{0}}%
\tau_{\eta^{\prime}}^{sk}\right] \nonumber\\
&  +\sum_{\eta^{\prime}}\int\frac{d\phi^{\prime}}{2\pi}\omega_{\eta\phi
,\eta^{\prime}\phi^{\prime}}^{a}\sin\left(  \phi^{\prime}-\phi\right)
\frac{v_{\eta^{\prime}}^{0}}{v_{\eta}^{0}}\tau_{\eta^{\prime}}^{L}.
\label{SBE-2-2}%
\end{align}

On the other hand, further analysis of Eq. (\ref{SBE-a-Born-1}) in general
cases of multiple Fermi circles is difficult, because we do not have simpler
general expression for the inter-band component of side-jump velocity.

\subsection{Decoupling condition for multiple Fermi circles in the SBE}

Although the $1/\tau^{||}$\ $\&$\ $1/\tau^{\perp}$ solution cannot be derived
out in general cases of isotropic 2DEG with multiple Fermi circles, it may
still be valid in some special multiple-Fermi-circles cases. In this
subsection we examine in what cases the $1/\tau^{||}$\ $\&$\ $1/\tau^{\perp}$
solution solves Boltzmann equations (\ref{SBE-n-origin}) and
(\ref{SBE-a-origin}) in isotropic multiple-Fermi-circle 2DEG. In the isotropic
multiple-Fermi--circle case the $1/\tau^{||}$\ $\&$\ $1/\tau^{\perp}$ solution
(\ref{complete}) takes the form \cite{Schliemann2003,Borunda2007}%
\begin{equation}
\frac{1}{\tau_{\eta}^{||}}=\sum_{\eta^{\prime}}\int\frac{d\phi^{\prime}}{2\pi
}\omega_{\eta\phi,\eta^{\prime}\phi^{\prime}}^{s}\left[  1-\frac
{v_{\eta^{\prime}}^{0}}{v_{\eta}^{0}}\cos\left(  \phi^{\prime}-\phi\right)
\right]  , \label{tau-p-multi}%
\end{equation}%
\begin{equation}
\frac{1}{\tau_{\eta}^{\perp}}=\sum_{\eta^{\prime}}\int\frac{d\phi^{\prime}%
}{2\pi}\omega_{\eta\phi,\eta^{\prime}\phi^{\prime}}^{a}\frac{v_{\eta^{\prime}%
}^{0}}{v_{\eta}^{0}}\sin\left(  \phi-\phi^{\prime}\right)  .
\label{tau-v-multi}%
\end{equation}
In this solution, $\tau_{\eta}^{\Vert\left(  \bot\right)  }$, thus $\tau
_{\eta}^{L\left(  sk\right)  }$ on different Fermi circles are decoupled,
however, they are coupled to each other in Eqs. (\ref{SBE-1-2}) and
(\ref{SBE-2-2}). Thereby, the $1/\tau^{||}$\ $\&$\ $1/\tau^{\perp}$ solution
is not expected to be suitable as long as the coupling between different Fermi
circles exists. While, the inter-band coupling vanishes when the following
decoupling condition i.e., when
\begin{equation}
\int d\phi^{\prime}\omega_{\eta\phi,\eta^{\prime}\phi^{\prime}}^{s}\cos\left(
\phi^{\prime}-\phi\right)  =0 \label{exist-s}%
\end{equation}
and%
\begin{equation}
\int d\phi^{\prime}\omega_{\eta\phi,\eta^{\prime}\phi^{\prime}}^{a}\sin\left(
\phi^{\prime}-\phi\right)  =0 \label{exist-a}%
\end{equation}
for $\eta^{\prime}\neq\eta$, is satisfied, then Eqs. (\ref{SBE-1-2}) and
(\ref{SBE-2-2}) reduce to Eq. (\ref{complete}) with (\ref{tau-p-multi}) and
(\ref{tau-v-multi}). Meanwhile, if the decoupling condition is satisfied, Eq.
(\ref{SBE-a-Born-1}) just yields $\tau_{\eta}^{sj}$ as $\tau_{\eta}^{L}$
calculated in the first Born order.

In the spin-polarized Rashba 2DEG with pointlike scalar impurities when both
Rashba bands are partially occupied, neither (\ref{exist-s}) nor
(\ref{exist-a}) holds when both Rashba bands are partially occupied. This can
be verified by noticing the form of Eqs. (\ref{scattering rate-2s}),
(\ref{scattering rate-3nG-1}), (\ref{scattering rate 4G}) and
(\ref{scattering rate 4nG}) in Sec. IV. It is therefore expected that the
$1/\tau^{||}$\ $\&$\ $1/\tau^{\perp}$ solution cannot provide fully correct
anomalous Hall conductivities for this system.

\subsection{Approximate strategy and scattering rates}

Usually the scattering chirality is weak \cite{SinitsynPRL2006} $1/\tau_{\eta
}^{||}\gg1/\tau_{\eta}^{\perp}$, thus the $1/\tau^{||}$\ $\&$\ $1/\tau^{\perp
}$ solution takes the form \cite{Sinitsyn2007,Borunda2007}
\begin{equation}
\tau_{\eta}^{L}=\tau_{\eta}^{||},\text{ \ \ }\tau_{\eta}^{sk}=\left(
\tau_{\eta}^{||}\right)  ^{2}/\tau_{\eta}^{\perp}, \label{usual}%
\end{equation}
with $\tau_{\eta}^{L}/\tau_{\eta}^{sk}\gg1$. Meanwhile, on the rhs of Eq.
(\ref{SBE-1-2}) the second term containing both $\omega^{a}$ and $\tau^{sk}$
may be disregarded. Eq. (\ref{SBE-1-2}) can thus be approximated by
\begin{equation}
1=\sum_{\eta^{\prime}}\int\frac{d\phi^{\prime}}{2\pi}\omega_{\eta\phi
,\eta^{\prime}\phi^{\prime}}^{s}\left[  \tau_{\eta}^{L}-\cos\left(
\phi^{\prime}-\phi\right)  \frac{v_{\eta^{\prime}}^{0}}{v_{\eta}^{0}}%
\tau_{\eta^{\prime}}^{L}\right]  , \label{SBE-1-3}%
\end{equation}
decoupled from Eq. (\ref{SBE-2-2}). Once $\tau_{\eta}^{L}$ is obtained from
Eq. (\ref{SBE-1-3}), it can be substituted into Eq. (\ref{SBE-2-2}) to get
$\tau_{\eta}^{sk}$. When the decoupling condition is satisfied, Eqs.
(\ref{SBE-1-3}) and (\ref{SBE-2-2}) just reduce to Eq. (\ref{usual}) with
(\ref{tau-p-multi}) and (\ref{tau-v-multi}).

Concrete expressions of $\omega_{\eta\phi,\eta^{\prime}\phi^{\prime}%
}^{s\left(  a\right)  }$ can be obtained by truncating the T-matrix expansion
\cite{Sinitsyn2007}. Randomly distributed identical scalar $\delta$-scatterers
are assumed for simplicity: $V\left(  \mathbf{r}\right)  =\sum_{i}V_{0}%
\delta\left(  \mathbf{r}-\mathbf{R}_{i}\right) $. In calculating the skew
scattering contributions, we take into account the non-Gaussian disorder
correlation $\left\langle V_{\mathbf{kk}^{\prime}}V_{\mathbf{k}^{\prime
}\mathbf{k}^{\prime\prime}}V_{\mathbf{k}^{\prime\prime}\mathbf{k}%
}\right\rangle _{c}=n_{im}V_{0}^{3}$ and $\left\langle V_{\mathbf{kk}^{\prime
}}V_{\mathbf{k}^{\prime}\mathbf{k}^{\prime\prime}}V_{\mathbf{k}^{\prime\prime
}\mathbf{k}^{\prime\prime\prime}}V_{\mathbf{k}^{\prime\prime\prime}\mathbf{k}%
}\right\rangle _{c}=n_{im}V_{0}^{4}$ other than the Gaussian correlation
$\left\langle V_{\mathbf{kk}^{\prime}}V_{\mathbf{k}^{\prime}\mathbf{k}%
}\right\rangle =n_{im}V_{0}^{2}$. Here $V_{\mathbf{kk}^{\prime}}$ is the
spin-independent part of the disorder matrix element, $n_{im}$ is the impurity
concentration, $\left\langle ..\right\rangle _{c}$ denotes the connected part
of disorder correlation. The non-Gaussian correlation in $o\left(
V^{4}\right)  $ usually yields a skew scattering contribution which is a
higher order quantity in terms of disorder strength relative to the
conventional skew scattering arising from the $o\left(  V^{3}\right)  $
non-Gaussian correlation and thus can be neglected \cite{Luttinger1958}.
However, if the conventional skew scattering contribution vanishes due to some
special band structures \cite{Borunda2007,Nunner2007}, the skew scattering
induced by the $o\left(  V^{4}\right)  $ non-Gaussian correlation plays an
important role in the weak scattering regime \cite{Kovalev2008,Kovalev2009}.

In the golden rule when calculated to $o\left(  V^{4}\right)  $, i.e., the
third Born order, the scattering rate is approximated as \cite{Sinitsyn2008}
(details in Appendix)
\begin{equation}
\omega_{l,l\prime}\simeq\omega_{l,l\prime}^{2s}+\omega_{l,l\prime}%
^{a-3nG}+\omega_{l,l\prime}^{a-4G}+\omega_{l,l\prime}^{a-4nG}%
.\label{total scattering rate}%
\end{equation}
In the second Born order $\omega_{l,l\prime}^{a-3nG}$ is related to
non-Gaussian disorder correlation, whereas in the third Born order both
Gaussian ($\omega_{l,l\prime}^{a-4G}$) and non-Gaussian ($\omega_{l,l\prime
}^{a-4nG}$) contributions are present. The three anti-symmetric scattering
rates on the rhs of Eq. (\ref{total scattering rate}) have distinct dependence
on the impurity concentration and scattering strength: $\omega_{l,l\prime
}^{a-3nG}\sim$ $n_{im}V_{0}^{3}$, $\omega_{l,l\prime}^{a-4nG}\sim n_{im}%
V_{0}^{4}$, $\omega_{l,l\prime}^{a-4G}\sim\left(  n_{im}V_{0}^{2}\right)
^{2}$. Because Eq. (\ref{SBE-2-2}) is linear with respect to $\tau^{sk}$ and
$\omega^{a}$, there exist three distinct skew scattering processes described
by $\tau_{\eta}^{sk-3nG}$, $\tau_{\eta}^{sk-4nG}$ and $\tau_{\eta}^{sk-4G}$
which correspond to $\omega_{l,l\prime}^{a-3nG}$, $\omega_{l,l\prime}^{a-4nG}$
and $\omega_{l,l\prime}^{a-4G}$, respectively:%
\begin{equation}
\tau_{\eta}^{sk}=\tau_{\eta}^{sk-3nG}+\tau_{\eta}^{sk-4nG}+\tau_{\eta}%
^{sk-4G}.\label{skew time}%
\end{equation}
Thus Eq. (\ref{SBE-2-2}) is decomposed into three independent linear systems%
\begin{align}
0 &  =\sum_{\eta^{\prime}}\int\frac{d\phi^{\prime}}{2\pi}\omega_{\eta\phi
,\eta^{\prime}\phi^{\prime}}^{2s}\left[  \tau_{\eta}^{sk-i}-\cos\left(
\phi^{\prime}-\phi\right)  \frac{v_{\eta^{\prime}}^{0}}{v_{\eta}^{0}}%
\tau_{\eta^{\prime}}^{sk-i}\right]  \nonumber\\
&  +\sum_{\eta^{\prime}}\int\frac{d\phi^{\prime}}{2\pi}\omega_{\eta\phi
,\eta^{\prime}\phi^{\prime}}^{a-i}\sin\left(  \phi^{\prime}-\phi\right)
\frac{v_{\eta^{\prime}}^{0}}{v_{\eta}^{0}}\tau_{\eta^{\prime}}^{L}%
,\label{SBE-2-3}%
\end{align}
here the superscript $i=3nG$, $4nG$ and $4G$ stand for distinct skew
scattering contributions. From Eq. (\ref{SBE-2-3}) one can see that
$\tau_{\eta}^{sk-3nG}\sim\left(  n_{im}V_{0}\right)  ^{-1}$, $\tau_{\eta
}^{sk-4nG}\sim n_{im}^{-1}$ and $\tau_{\eta}^{sk-4G}\sim\left(  n_{im}%
V_{0}\right)  ^{0}$. Based on Eq. (\ref{skew time}), Eq. (\ref{DF}) can be
decomposed into $g_{\eta}=g_{\eta}^{L}+\sum_{i}g_{\eta}^{sk-i}$ where
$g_{\eta}^{L}=\left(  -\partial_{\epsilon}f^{0}\right)  \mathbf{F}%
\cdot\mathbf{v}_{\eta}^{0}\left(  \epsilon,\phi\right)  \tau_{\eta}^{L}\left(
\epsilon\right)  $ denotes the DF responsible for the longitudinal transport
and $g_{\eta}^{sk-i}=\left(  -\partial_{\epsilon}f^{0}\right)  \left(
\mathbf{\hat{z}\times F}\right)  \cdot\mathbf{v}_{\eta}^{0}\left(
\epsilon,\phi\right)  \tau_{\eta}^{sk-i}\left(  \epsilon\right)  $ denotes the
skew scattering contribution to the out-of-equilibrium DF.

$\tau_{\eta}^{sk-4G}$ is independent of the disorder strength and impurity
concentration, so does the resulting anomalous Hall conductivity. Thereby this
Gaussian skew scattering is termed \textquotedblleft intrinsic skew
scattering\textquotedblright\ by Sinitsyn \cite{Sinitsyn2008}. While, the
anomalous Hall conductivity arising from the non-Gaussian skew scattering in
the third Born order is inversely proportional to the impurity concentration
but independent of the disorder strength. This character lies between the
conventional skew scattering and side-jump mechanisms, thus this mechanism is
termed \textquotedblleft hybrid skew scattering\textquotedblright%
\ \cite{Kovalev2008,Kovalev2009}.

\section{Calculations}

\subsection{Model}

We consider the spin-polarized Rashba 2DEG
\begin{equation}
\hat{H}=\frac{\mathbf{\hat{p}}^{2}}{2m}+\frac{\alpha}{\hbar}\mathbf{\hat
{\sigma}}\cdot\left(  \mathbf{\hat{p}}\times\mathbf{\hat{z}}\right)
-J_{ex}\hat{\sigma}_{z}+V\left(  \mathbf{r}\right)  ,
\end{equation}
where $V\left(  \mathbf{r}\right)  $ is the disorder potential introduced in
Sec. III. D, $m$ is the in-plane effective mass of the conduction electron,
$\mathbf{\hat{p}=\hbar\hat{k}}$ the 2D momentum, $\mathbf{\hat{\sigma}%
=}\left(  \hat{\sigma}_{x},\hat{\sigma}_{y},\hat{\sigma}_{z}\right)  $ are the
Pauli matrices, $\alpha$ the Rashba spin-orbital coupling coefficient,
$J_{ex}$ the exchange field. Inner Eigenstates of the pure system read
\begin{equation}
|u_{\mathbf{k}}^{\eta}\rangle=\frac{1}{\sqrt{2}}%
\genfrac{[}{]}{0pt}{}{\sqrt{1-\eta\cos\theta}}{-i\eta\exp\left(  i\phi\right)
\sqrt{1+\eta\cos\theta}}%
,
\end{equation}
where $\eta=\pm,\cos\theta=J_{ex}/\Delta_{k},\Delta_{k}=\sqrt{\alpha^{2}%
k^{2}+J_{ex}^{2}}$, $\sin\theta=\alpha k/\Delta_{k},\tan\phi_{k}=k_{y}/k_{x}$.
In the present paper we only consider the case when $\epsilon_{F}>J_{ex}%
$\textbf{,} i.e., both Rashba bands are partially occupied. For any energy
$\epsilon>J_{ex}$ there are two iso-energy circles corresponding to the two
Rashba bands: $k_{\eta}^{2}\left(  \epsilon\right)  =\frac{2m}{\hbar^{2}%
}\left(  \epsilon-\eta\Delta_{\eta}\left(  \epsilon\right)  \right)  $ where
we define $\epsilon_{R}=m\left(  \frac{\alpha}{\hbar}\right)  ^{2}$ and
$\Delta_{\eta}\left(  \epsilon\right)  =\sqrt{\epsilon_{R}^{2}+J_{ex}%
^{2}+2\epsilon_{R}\epsilon}-\eta\epsilon_{R}$. The DOS in the $\eta$ band is
$N_{\eta}\left(  \epsilon\right)  =N_{0}\frac{\Delta_{\eta}\left(
\epsilon\right)  }{\Delta_{\eta}\left(  \epsilon\right)  +\eta\epsilon_{R}}$
with $N_{0}=\frac{m}{2\pi\hbar^{2}}$. The intra-band and inter-band
energy-integrated scattering rates in the lowest Born order read $\omega
_{\eta\phi,\eta^{\prime}\phi^{\prime}}^{2s}\left(  \epsilon\right)
=\frac{N_{\eta^{\prime}}\left(  \epsilon\right)  }{\tau N_{0}}\left\vert
\langle u_{\mathbf{k}_{\eta}\left(  \epsilon\right)  }^{\eta}|u_{\mathbf{k}%
_{\eta^{\prime}}^{\prime}\left(  \epsilon\right)  }^{\eta^{\prime}}%
\rangle\right\vert ^{2}$, where%
\begin{gather}
\left\vert \langle u_{\mathbf{k}_{\eta}\left(  \epsilon\right)  }^{\eta
}|u_{\mathbf{k}_{\eta^{\prime}}^{\prime}\left(  \epsilon\right)  }%
^{\eta^{\prime}}\rangle\right\vert ^{2}=\frac{1}{2}\left[  1+\eta\eta^{\prime
}\frac{J_{ex}^{2}}{\Delta_{\eta}\left(  \epsilon\right)  \Delta_{\eta^{\prime
}}\left(  \epsilon\right)  }\right.  \nonumber\\
\left.  +\eta\eta^{\prime}\cos\left(  \phi^{\prime}-\phi\right)  \frac
{\alpha^{2}k_{\eta}\left(  \epsilon\right)  k_{\eta^{\prime}}\left(
\epsilon\right)  }{\Delta_{\eta}\left(  \epsilon\right)  \Delta_{\eta^{\prime
}}\left(  \epsilon\right)  }\right]  ,\label{scattering rate-2s}%
\end{gather}
with $\tau=\left(  2\pi n_{im}V_{0}^{2}N_{0}/\hbar\right)  ^{-1}$.

\subsection{Coordinate-shift effects}

\subsubsection{\textbf{Intra-band and inter-band side-jump velocities}}

The side-jump velocity $\mathbf{v}_{\eta\mathbf{k}}^{sj}=\sum_{\eta^{\prime
}\mathbf{k}^{\prime}}\omega_{\eta^{\prime}\mathbf{k}^{\prime},\eta\mathbf{k}%
}^{2s}\delta\mathbf{r}_{\eta^{\prime}\mathbf{k}^{\prime},\eta\mathbf{k}}$ can
be further decomposed into intra-band and inter-band components:%
\begin{equation}
\mathbf{v}_{\eta\mathbf{k}}^{sj}=\left(  \mathbf{v}_{\eta\mathbf{k}}%
^{sj}\right)  ^{intra}+\left(  \mathbf{v}_{\eta\mathbf{k}}^{sj}\right)
^{inter}, \label{sj velocity-total}%
\end{equation}
where the intra-band and inter-band\ side-jump velocities are
\begin{equation}
\left(  \mathbf{v}_{\eta\mathbf{k}}^{sj}\right)  ^{intra}=\sum_{\mathbf{k}%
^{\prime}}\omega_{\eta\mathbf{k}^{\prime},\eta\mathbf{k}}^{2s}\delta
\mathbf{r}_{\eta\mathbf{k}^{\prime},\eta\mathbf{k}}%
\end{equation}
and
\begin{equation}
\left(  \mathbf{v}_{\eta\mathbf{k}}^{sj}\right)  ^{inter}=\sum_{\mathbf{k}%
^{\prime}}\omega_{-\eta\mathbf{k}^{\prime},\eta\mathbf{k}}^{2s}\delta
\mathbf{r}_{-\eta\mathbf{k}^{\prime},\eta\mathbf{k}},
\end{equation}
respectively. The intra-band (inter-band) component originates from the
coordinate-shift process with initial and final states on the same Rashba band
(different Rashba bands). Straightforward evaluations lead to $\left(
\mathbf{v}_{\eta}^{sj}\left(  \epsilon,\phi\right)  \right)  _{x}%
^{intra}=k_{\eta}\left(  \epsilon\right)  \sin\phi\frac{\eta\alpha^{2}J_{ex}%
}{2\Delta_{\eta}^{3}\left(  \epsilon\right)  }\frac{N_{\eta}\left(
\epsilon\right)  }{\tau N_{0}}$. On the other hand, this intra-band result can
be obtained directly from the expression for the intra-band coordinate-shift
\cite{Sinitsyn2007} $\delta\mathbf{r}_{\eta\mathbf{k}^{\prime},\eta\mathbf{k}%
}=\Omega_{\eta k}\mathbf{\hat{z}\times}\left(  \mathbf{k}^{\prime}%
-\mathbf{k}\right)  /\left\vert \langle u_{\mathbf{k}}^{\eta}|u_{\mathbf{k}%
^{\prime}}^{\eta}\rangle\right\vert ^{2}$. Here $\Omega_{\eta\mathbf{k}}$ is
the z-component of Berry-curvature of Rashba band $\eta$ at momentum
$\mathbf{k}$. Thereby, the intra-band side-jump velocity also reads%
\begin{equation}
\left(  \mathbf{v}_{\eta}^{sj}\left(  \epsilon,\phi\right)  \right)
^{intra}=\frac{N_{\eta}\left(  \epsilon\right)  }{\tau N_{0}}\Omega_{\eta
}\left(  \epsilon\right)  \mathbf{k}_{\eta}\left(  \epsilon\right)
\times\mathbf{\hat{z}},
\end{equation}
with $\Omega_{\eta}\left(  \epsilon\right)  =\eta\frac{\alpha^{2}J_{ex}%
}{2\Delta_{\eta}^{3}\left(  \epsilon\right)  }$ the Berry-curvature on the
iso-energy circle in momentum-space. It is known that coordinate-shift is an
inter-band coherence effect induced by the scattering potential, this is just
reflected in the above formula which contains the Berry-curvature implying
that the virtual vertical inter-band transition contributes to the side-jump
velocity even in the intra-band scattering process. In the 2DEG with single
Fermi circle such as 2D massive Dirac model \cite{Sinitsyn2007}, only the
intra-band elastic scattering process exists. However, in the present system
with two Fermi circles and short-range disorder, coordinate-shift effects also
occur in the inter-band elastic scattering process. While the inter-band
side-jump velocity is absent in the limit of smooth disorder potential
\cite{Sinitsyn2005}, in the considered case it is obtained as%
\begin{equation}
\left(  \mathbf{v}_{\eta}^{sj}\left(  \epsilon,\phi\right)  \right)
_{x}^{inter}=\frac{-\eta\epsilon_{R}}{\Delta_{-\eta}\left(  \epsilon\right)
}\left(  \mathbf{v}_{\eta}^{sj}\left(  \epsilon,\phi\right)  \right)
_{x}^{intra}.
\end{equation}
For the inner (outer) Rashba band the inter-band side-jump velocity is
anti-parallel (parallel) to the intra-band one and thus decreases (enhances)
the total side-jump velocity. For the inner Rashba band, the magnitude of the
inter-band side-jump velocity is always smaller than that of the intra-band
one. However, for the outer Rashba band, the inter-band side-jump velocity
exceeds the intra-band one when%
\begin{equation}
\epsilon<\epsilon_{c}\equiv\frac{3\epsilon_{R}^{2}-J_{ex}^{2}}{2\epsilon_{R}}.
\end{equation}
For fixed $J_{ex}$ $\left(  \epsilon_{R}\right)  $, $\epsilon_{c}$ increases
(decreases) with increasing $\epsilon_{R}$ $\left(  J_{ex}\right)  $. One
finds that the inter-band coordinate-shift contribution is more important for
larger spin-orbital coupling, this can be clearly seen in figure 1, which
shows the ratio $\left(  v_{\eta}^{sj}\left(  \epsilon\right)  \right)
^{inter}/v_{\eta}^{sj}\left(  \epsilon\right)  =\epsilon_{R}/\sqrt
{\epsilon_{R}^{2}+J_{ex}^{2}+2\epsilon_{R}\epsilon}$ with $v_{\eta}%
^{sj}\left(  \epsilon\right)  \equiv\left\vert \mathbf{v}_{\eta}^{sj}\left(
\epsilon,\phi\right)  \right\vert $.

\begin{figure}[ptbh]
\includegraphics[width=0.35\textwidth]{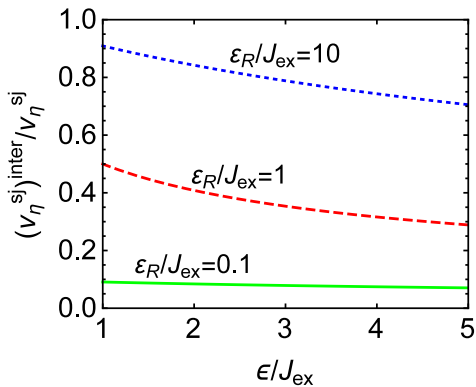} \caption{The inter-band contribution to the
side-jump velocity for various model parameters. When the spin-orbital coupling is stronger, the inter-band
contribution becomes more important.}%
\label{fig1}%
\end{figure}

The vector form of the total side-jump velocity reads%
\begin{equation}
\mathbf{v}_{\eta}^{sj}\left(  \epsilon,\phi\right)  =\eta\mathbf{k}_{\eta
}\left(  \epsilon\right)  \times\mathbf{\hat{z}}\frac{J_{ex}\alpha^{2}%
}{2\Delta_{\eta}\left(  \epsilon\right)  \left(  J_{ex}^{2}+2\epsilon
_{R}\epsilon\right)  }\frac{1}{\tau}. \label{sj velocity-total-1}%
\end{equation}
The key is that $\mathbf{v}^{sj}$ is orthogonal to $\mathbf{k}_{\eta}\left(
\epsilon\right)  $, and the side-jump velocities of the inner and outer Rashba
bands have opposite chirality.

\subsubsection{\textbf{Solution of the anomalous DF}}

Substituting Eq. (\ref{sj velocity-total-1}) into (\ref{SBE-a-Born-1}) yields
the $2\times2$ linear system ($\eta=\pm$) determining $\tau_{\eta}^{sj}\left(
\epsilon\right)  $:
\begin{equation}
1=\sum_{\eta^{\prime}}\int\frac{d\phi^{\prime}}{2\pi}\omega_{\eta\phi
,\eta^{\prime}\phi^{\prime}}^{2s}\left[  \tau_{\eta}^{sj}-\frac{\eta^{\prime
}k_{\eta^{\prime}}\Delta_{\eta}}{\eta k_{\eta}\Delta_{\eta^{\prime}}}%
\cos\left(  \phi^{\prime}-\phi\right)  \tau_{\eta^{\prime}}^{sj}\right]  .
\label{tau-sj}%
\end{equation}
The factor $\eta^{\prime}/\eta$ in the second term inside the square bracket
just reflects different chirality of side-jump velocity on the two Rashba
bands. It follows that%
\begin{equation}
\tau_{\eta}^{sj}\left(  \epsilon\right)  =\tau\frac{J_{ex}^{2}+2\epsilon
_{R}\epsilon}{J_{ex}^{2}+\epsilon_{R}\epsilon}, \label{tau-sj-result}%
\end{equation}
which does not depend on the band index $\eta$. The anomalous part of the
out-of-equilibrium DF is then obtained as%
\begin{equation}
g_{\eta}^{adis}\left(  \epsilon\right)  =\eta\left(  \mathbf{\hat{z}\times
F}\right)  \cdot\mathbf{k}_{\eta}\left(  \epsilon\right)  \frac{J_{ex}%
\alpha^{2}}{2\Delta_{\eta}\left(  \epsilon\right)  \left(  J_{ex}^{2}%
+\epsilon_{R}\epsilon\right)  }\frac{\partial f^{0}}{\partial\epsilon},
\end{equation}
which is different from the one in the limit of smooth disorder potential in
which only the intra-band small angle electron-impurity scattering dominates
\cite{Sinitsyn2005}.

\subsection{DF responsible for the skew scattering}

The derivation of the needed scattering rates in this section is provided in
the Appendix.

\subsubsection{\textbf{Longitudinal nonequilibrium DF}}

The correct calculation of $\tau_{\eta}^{L}$ is very crucial in order to
obtain the skew scattering contributions to AHE via Eq. (\ref{SBE-2-3}). The
appropriate starting point is just the $2\times2$ linear system ($\eta=\pm$)
Eq. (\ref{SBE-1-3}). Substituting corresponding quantities, the longitudinal
transport time is obtained as%
\begin{equation}
\tau_{\eta}^{L}\left(  \epsilon\right)  =\tau\frac{N_{\eta}\left(
\epsilon\right)  }{N_{0}}. \label{longitudinal time}%
\end{equation}
It follows that $g_{\eta}^{L}\left(  \epsilon,\vartheta\left(  \mathbf{k}%
_{\eta}\left(  \epsilon\right)  \right)  \right)  =\left(  -\partial
_{\epsilon}f^{0}\right)  \mathbf{F}\cdot\frac{\hbar\mathbf{k}_{\eta}\left(
\epsilon\right)  }{m}\tau$ since $\mathbf{v}_{\eta}^{0}\left(  \epsilon
,\phi\right)  =\frac{\hbar\mathbf{k}_{\eta}\left(  \epsilon\right)  }{m}%
\frac{N_{0}}{N_{\eta}\left(  \epsilon\right)  }$. This solution has the same
form as that in a nonmagnetic Rashba 2DEG \cite{Xiao2016FOP}.

\subsubsection{\textbf{Conventional skew scattering}}

Due to $N_{-\eta}\left(  \epsilon\right)  /N_{\eta}\left(  \epsilon\right)
=\Delta_{-\eta}\left(  \epsilon\right)  /\Delta_{\eta}\left(  \epsilon\right)
$, the anti-symmetric part of the energy-integrated third-order Gaussian
scattering rate vanishes%
\begin{equation}
\omega_{\eta\phi,\eta^{\prime}\phi^{\prime}}^{a-3nG}\left(  \epsilon\right)
\sim\sin\left(  \phi^{\prime}-\phi\right)  \sum_{\eta^{\prime\prime}}%
\eta^{\prime\prime}\frac{N_{\eta^{\prime\prime}}\left(  \epsilon\right)
}{\Delta_{\eta^{\prime\prime}}\left(  \epsilon\right)  }=0,
\label{scattering rate-3nG-1}%
\end{equation}
then $\tau_{\eta}^{sk-3nG}\left(  \epsilon\right)  =0$ and no corresponding
nonequilibrium response arises when both Rashba bands are partially occupied.
We emphasize that the absence of conventional skew scattering stems directly
from the vanishing $\omega_{\eta\phi,\eta^{\prime}\phi^{\prime}}^{a-3nG}$.
Thus the previous SBE calculation \cite{Borunda2007} in which the
multiple-Fermi-circle SBE was solved by using the $1/\tau^{||}$\ $\&$%
\ $1/\tau^{\perp}$ solution also produced this result (via Eq. (\ref{comp-nG})).

\subsubsection{\textbf{Intrinsic skew scattering}}

The anti-symmetric part of the energy-integrated fourth-order Gaussian
scattering rate reads
\begin{align}
\omega_{\eta\phi,\eta^{\prime}\phi^{\prime}}^{a-4G}\left(  \epsilon\right)
&  =\frac{\hbar N_{\eta^{\prime}}\left(  \epsilon\right)  }{4N_{0}\tau^{2}%
}\eta\eta^{\prime}\frac{J_{ex}}{J_{ex}^{2}+2\epsilon_{R}\epsilon}\frac
{\alpha^{2}k_{\eta}\left(  \epsilon\right)  k_{\eta^{\prime}}\left(
\epsilon\right)  }{\Delta_{\eta}\left(  \epsilon\right)  \Delta_{\eta^{\prime
}}\left(  \epsilon\right)  }\nonumber\\
&  \times\sin\left(  \phi^{\prime}-\phi\right)  . \label{scattering rate 4G}%
\end{align}
Substituting the expression of $\tau_{\eta}^{L}$ and scattering rates into the
SBE (\ref{SBE-2-3}) for $i=4G$ yields the following $2\times2$ linear system
for $\tau_{\pm}^{sk-4G}$
\begin{gather}
\left[  \frac{J_{ex}^{2}}{\Delta_{\eta}^{2}}+\frac{1}{4}\frac{\alpha
^{2}k_{\eta}^{2}}{\Delta_{\eta}^{2}}+\frac{N_{-\eta}}{2N_{\eta}}\left(
1-\frac{J_{ex}^{2}}{\Delta_{+}\Delta_{-}}\right)  \right]  \tau_{\eta}%
^{sk-4G}\nonumber\\
+\frac{1}{4}\frac{\alpha^{2}k_{-\eta}^{2}}{\Delta_{+}\Delta_{-}}\tau_{-\eta
}^{sk-4G}=\frac{\eta J_{ex}\epsilon_{R}}{J_{ex}^{2}+2\epsilon_{R}\epsilon
}\frac{\hbar}{2\Delta_{\eta}}. \label{linear system-4G}%
\end{gather}
Then we obtain the transverse transport time induced by the 4th order Gaussian
skew scattering as%
\begin{equation}
\tau_{\eta}^{sk-4G}\left(  \epsilon\right)  =\frac{\eta J_{ex}\epsilon
_{R}\hbar}{2\left(  J_{ex}^{2}+\epsilon_{R}\epsilon\right)  }\frac{1}%
{\sqrt{\epsilon_{R}^{2}+J_{ex}^{2}+2\epsilon_{R}\epsilon}}.
\label{sk-time-intrinsic}%
\end{equation}
This quantity does not depend on the impurity concentration and scattering
strength, just the characteristic of intrinsic skew scattering. The
corresponding part of the nonequilibrium DF is given by%
\begin{equation}
g_{\eta}^{sk-4G}\left(  \epsilon\right)  =-g_{\eta}^{adis}\left(
\epsilon\right)  .
\end{equation}
We see that, when both Rashba bands are partially occupied, the transport
contributions from the anomalous distribution related to the coordinate-shift
and that from the intrinsic skew scattering cancel each other exactly.

\subsubsection{\textbf{Hybrid skew scattering}}

The anti-symmetric part of the energy-integrated fourth-order non-Gaussian
scattering rate is obtained to be (assume $\epsilon_{R}\leq J_{ex}$)%
\begin{equation}
\omega_{\eta\phi,\eta^{\prime}\phi^{\prime}}^{a-4nG}\left(  \epsilon\right)
=\frac{\hbar N_{\eta}\left(  \epsilon\right)  }{4n_{im}\tau^{2}}\eta
\eta^{\prime}\frac{\alpha^{2}k_{\eta}\left(  \epsilon\right)  k_{\eta^{\prime
}}\left(  \epsilon\right)  }{\Delta_{\eta}\left(  \epsilon\right)
\Delta_{\eta^{\prime}}\left(  \epsilon\right)  }I\left(  \epsilon\right)
\sin\left(  \phi^{\prime}-\phi\right)  , \label{scattering rate 4nG}%
\end{equation}
where $I\left(  \epsilon\right)  $ is given by Eq. (\ref{I}). Substituting Eq.
(\ref{longitudinal time}) and relevant scattering rates into Eq.
(\ref{SBE-2-3}) for $i=4nG$ yields two coupled equations (not shown)
determining $\tau_{\pm}^{sk-4nG}$. Comparing these equations to Eq.
(\ref{linear system-4G}), one can find directly that
\begin{equation}
\tau_{\eta}^{sk-4nG}\left(  \epsilon\right)  =\tau_{\eta}^{sk-4G}\left(
\epsilon\right)  S\left(  \epsilon\right)  ,
\end{equation}
where $S\left(  \epsilon\right)  =I\left(  \epsilon\right)  \frac{N_{0}%
}{n_{im}}\frac{J_{ex}^{2}+2\epsilon_{R}\epsilon}{J_{ex}}$ is band-independent.

\subsection{Results for AHE}

The anomalous Hall conductivity in the SBE framework is expressed as
\cite{Sinitsyn2007} $\sigma_{yx}=\sigma_{yx}^{in}+\sigma_{yx}^{sj}+\sigma
_{yx}^{adis}+\sum_{i}\sigma_{yx}^{sk-i}$, where $\sigma_{yx}^{sj}=e\sum
_{l}\left(  v_{l}^{sj}\right)  _{y}g_{\eta}^{L}/\left(  \mathbf{E}\right)
_{x}$ and $\sigma_{yx}^{adis\left(  sk-i\right)  }=e\sum_{l}\left(  v_{l}%
^{0}\right)  _{y}g_{\eta}^{adis\left(  sk-i\right)  }/\left(  \mathbf{E}%
\right)  _{x}$ represent the Hall conductivities due to the side-jump velocity
and anomalous DF (skew scattering), respectively. $\sigma_{yx}^{in}$ is the
intrinsic contribution determined by the momentum-space Berry-curvature
\cite{Niu2010}. When both Rashba bands are partially occupied, $g_{\eta
}^{adis}=-g_{\eta}^{sk-4G}$, $g_{\eta}^{sk-3nG}=0$ and $\sigma_{yx}%
^{adis}=\sigma_{yx}^{sj}=-\sigma_{yx}^{in}=\frac{e^{2}}{2\pi\hbar}%
J_{ex}\epsilon_{R}/\left(  J_{ex}^{2}+2\epsilon_{R}\epsilon_{F}\right)  $,
thus $\sigma_{yx}^{in}+\sigma_{yx}^{sj}+\sigma_{yx}^{adis}+\sigma_{yx}%
^{sk-4G}+\sigma_{yx}^{sk-3nG}=0$. This\ is just the result obtained by the
Kubo-Streda method in spin-$s_{z}$ representation \cite{Nunner2007}:
$\sigma_{yx}^{I\left(  a\right)  ,b}+\sigma_{yx}^{I\left(  a\right)  ,l}=0$,
$\sigma_{yx}^{sk-3nG}=0$, $\sigma_{yx}^{II}=0$. Here $\sigma_{yx}^{I\left(
a\right)  ,b}=\sigma_{yx}^{in}-\sigma_{yx}^{II}$ denotes the bare bubble
contribution with $\sigma_{yx}^{II}$ representing Streda's Fermi sea term, and
the ladder vertex correction \cite{Sinitsyn2007} $\sigma_{yx}^{I\left(
a\right)  ,l}=\sigma_{yx}^{sj}+\sigma_{yx}^{adis}+\sigma_{yx}^{sk-4G}$. The
hybrid skew scattering dominates when both Rashba bands are partially occupied
\cite{Kovalev2008}, $\sigma_{yx}=\sigma_{yx}^{sk-4nG}$ and
\begin{equation}
\sigma_{yx}^{sk-4nG}=-\frac{e^{2}}{2\pi\hbar}\frac{N_{0}\epsilon_{R}}{n_{im}%
}I\left(  \epsilon_{F}\right)  . \label{AHC}%
\end{equation}
With Eq. (\ref{I}), this formula just reproduces the result obtained by the
multi-band Keldysh formalism \cite{Kovalev2008,note}.

\section{Discussion and conclusion}

Let us start this discussion section with a remark on the notable
difference between the coordinate-shift contribution to the AHE in the
presence of pointlike impurities and that in the limit of smooth disorder
potential analyzed by Sinitsyn et al. \cite{Sinitsyn2005}. In the latter case
the coordinate-shift contribution ($\sigma_{yx}^{sj}+\sigma_{yx}^{adis}$) has
the same sign as the intrinsic one for sufficiently large Fermi energy, while
in the former case the coordinate-shift contribution has the opposite sign as
and is two times larger in magnitude than the intrinsic one.

Now we discuss the difference between the $1/\tau^{||}$\ $\&$\ $1/\tau^{\perp
}$ solution and our solution explicitly. For $\epsilon>J_{ex}$ in the
$1/\tau^{||}$\ $\&$\ $1/\tau^{\perp}$ solution one obtains \cite{note-1} the
longitudinal transport time $\tilde{\tau}_{\eta}^{L}=\tau_{\eta}^{||}$
different from $\tau_{\eta}^{L}$ given by Eq. (\ref{longitudinal time}):
\begin{align}
\frac{\tilde{\tau}_{\eta}^{L}}{\tau_{\eta}^{L}}  &  =\left(  \frac{N_{0}%
}{N_{\eta}}\right)  ^{2}\left[  \frac{J_{ex}^{2}}{\Delta_{\eta}^{2}}%
+\frac{\alpha^{2}k_{\eta}^{2}}{4\Delta_{\eta}^{2}}+\frac{\alpha^{2}k_{-\eta
}^{2}}{4\Delta_{\eta}\Delta_{-\eta}}\right. \nonumber\\
&  \left.  +\frac{N_{-\eta}}{2N_{\eta}}\left(  1-\frac{J_{ex}^{2}}%
{\Delta_{\eta}\Delta_{-\eta}}\right)  \right]  ^{-1}\label{comp-L}\\
&  =\frac{\left(  \epsilon_{R}^{2}+J_{ex}^{2}+2\epsilon_{R}\epsilon\right)
\left(  J_{ex}^{2}+2\epsilon_{R}\epsilon\right)  }{\left(  J_{ex}%
^{2}+2\epsilon_{R}\epsilon\right)  ^{2}-\eta\epsilon_{R}^{2}\epsilon\left(
\sqrt{\epsilon_{R}^{2}+J_{ex}^{2}+2\epsilon_{R}\epsilon}-\eta\epsilon
_{R}\right)  }.\nonumber
\end{align}
For the skew scattering, the $1/\tau^{||}$\ $\&$\ $1/\tau^{\perp}$ solution
\begin{equation}
\tilde{\tau}_{\eta}^{sk-i}=\left(  \tilde{\tau}_{\eta}^{L}\right)  ^{2}%
\sum_{\eta^{\prime}}\int\frac{d\phi^{\prime}}{2\pi}\omega_{\eta\phi
,\eta^{\prime}\phi^{\prime}}^{a-i}\frac{v_{\eta^{\prime}}^{0}}{v_{\eta}^{0}%
}\sin\left(  \phi-\phi^{\prime}\right)  \label{comp-nG}%
\end{equation}
yields $\tilde{\tau}_{\eta}^{sk-3nG}=0$ and
\begin{equation}
\frac{\tilde{\tau}_{\eta}^{sk-4G}\left(  \epsilon\right)  }{\tau_{\eta
}^{sk-4G}\left(  \epsilon\right)  }=\left(  \frac{\tilde{\tau}_{\eta}^{L}%
}{\tau}\frac{J_{ex}^{2}+\epsilon_{R}\epsilon}{J_{ex}^{2}+2\epsilon_{R}%
\epsilon}\right)  ^{2}. \label{comp-sk}%
\end{equation}
One can verify that $\omega_{\eta\phi,\eta^{\prime}\phi^{\prime}}%
^{a-4nG}\left(  \epsilon\right)  /\omega_{\eta\phi,\eta^{\prime}\phi^{\prime}%
}^{a-4G}\left(  \epsilon\right)  =S\left(  \epsilon\right)  $, thus
$\tilde{\tau}_{\eta}^{sk-4nG}\left(  \epsilon\right)  /\tilde{\tau}_{\eta
}^{sk-4G}\left(  \epsilon\right)  =S\left(  \epsilon\right)  $. On the other
hand, we have obtained $\tau_{\eta}^{sk-4nG}\left(  \epsilon\right)
/\tau_{\eta}^{sk-4G}\left(  \epsilon\right)  =S\left(  \epsilon\right)  $,
then%
\begin{equation}
\frac{\tilde{\tau}_{\eta}^{sk-4G}\left(  \epsilon\right)  }{\tau_{\eta
}^{sk-4G}\left(  \epsilon\right)  }=\frac{\tilde{\tau}_{\eta}^{sk-4nG}\left(
\epsilon\right)  }{\tau_{\eta}^{sk-4nG}\left(  \epsilon\right)  }=\frac
{\tilde{\tau}_{\eta}^{sk}\left(  \epsilon\right)  }{\tau_{\eta}^{sk}\left(
\epsilon\right)  }. \label{comp-sk-1}%
\end{equation}
Here $\tilde{\tau}_{\eta}^{sk}=\tilde{\tau}_{\eta}^{sk-3nG}+\tilde{\tau}%
_{\eta}^{sk-4G}+\tilde{\tau}_{\eta}^{sk-4nG}$. The comparisons of the
$1/\tau^{||}$\ $\&$\ $1/\tau^{\perp}$ solution and the present transport
relaxation time solution are shown clearly in figure 2. Figure 2(a) and 2(b)
show the ratio $\tilde{\tau}_{+}^{L}/\tau_{+}^{L}$ and $\tilde{\tau}_{-}%
^{L}/\tau_{-}^{L}$, respectively. When $\epsilon=J_{ex}$, Eq. (\ref{comp-L})
tells $\tilde{\tau}_{-}^{L}/\tau_{-}^{L}=1$. In the limit of large Fermi
energy, $\tilde{\tau}_{\eta}^{L}\simeq\tau_{\eta}^{L}\simeq\tau$; this
tendency is clear in figure 2(a) and will be clear in figure 2(b) if the
curves are plotted up to $\epsilon/J_{ex}\simeq50$ (not shown). Figure 2(c)
and 2(d) show the ratio $\tilde{\tau}_{+}^{sk}/\tau_{+}^{sk}$ and $\tilde
{\tau}_{-}^{sk}/\tau_{-}^{sk}$, respectively. Because in calculating the
hybrid skew scattering we assume $\epsilon_{R}\leq J_{ex}$, here we follow
this assumption. For the inner Rashba band, the deviation of $\tilde{\tau}%
_{+}^{sk}$ from $\tau_{+}^{sk}$ increases monotonically with increasing
$\epsilon_{R}/J_{ex}$ in the regime $\epsilon_{R}\leq J_{ex}$. In the limit of
large Fermi energy, $\tilde{\tau}_{\eta}^{sk}/\tau_{\eta}^{sk}\rightarrow1/4$;
this limiting behavior can be visible if the curves in figure 2(c) and 2(d)
are plotted up to $\epsilon/J_{ex}\simeq50$ (not shown). Whereas when
$\epsilon=J_{ex}$, Eq. (\ref{comp-sk}) tells $\tilde{\tau}_{-}^{sk}/\tau
_{-}^{sk}=1$.

\begin{figure}[ptbh]
\includegraphics[width=0.5\textwidth]{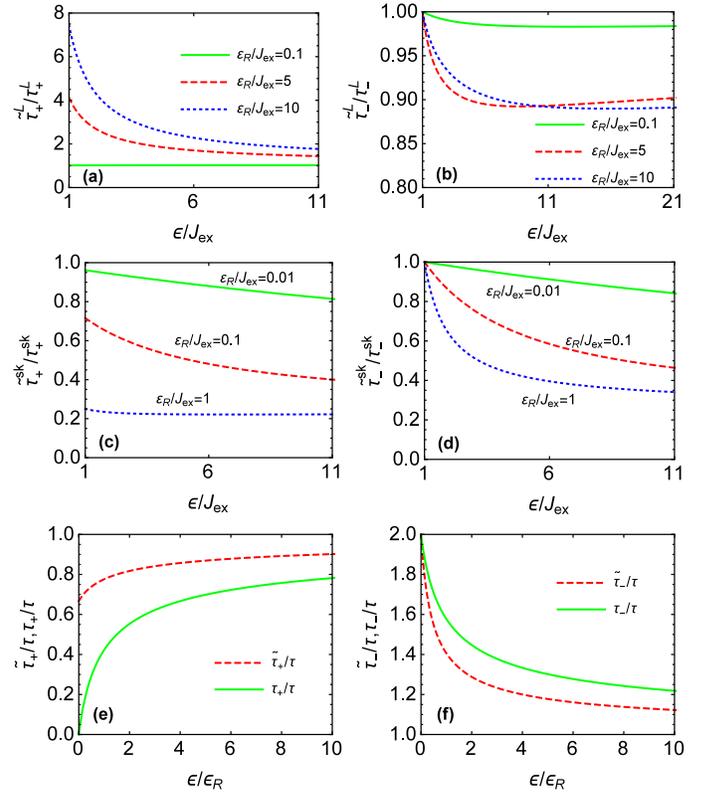} \caption{The
comparison of the $1/\tau^{||}$\ $\&$\ $1/\tau^{\perp}$ solution and our
solution. (a) and (b) for the longitudinal transport time, $\tilde{\tau}_{+}^{L}/\tau_{+}^{L}$ and $\tilde{\tau}_{-}^{L}/\tau_{-}^{L}$, respectively. (c) and (d) for the transport time responsible for skew-scattering-induced Hall transport, $\tilde{\tau}_{+}^{sk}/\tau_{+}^{sk}$ and $\tilde{\tau}_{-}^{sk}/\tau_{-}^{sk}$, respectively.
(e) and (f), comparison in the case of a nonmagnetic Rashba 2DEG.}%
\label{fig2}%
\end{figure}

To illustrate more transparently the difference between the $1/\tau^{||}%
$\ $\&$\ $1/\tau^{\perp}$ solution and the solution of Eqs. (\ref{SBE-1-3})
and (\ref{SBE-2-3}) when the decoupling condition is not satisfied, we also
consider the nonmagnetic Rashba 2DEG with pointlike potential scatterers as
the simplest example \cite{Xiao2016FOP} in figure 2(e) and 2(f). In this
isotropic 2DEG without AHE, the lowest Born approximation is sufficient for
calculating the scattering rate, thus $1/\tau_{\eta}^{\bot}=0$. When both
Rashba bands are partially occupied, the decoupling condition (\ref{exist-s})
is not satisfied and $\tilde{\tau}_{\eta}^{L}=\tau_{\eta}^{||}$ given by Eq.
(\ref{tau-p-multi}) differs from the solution of Eq. (\ref{SBE-1-3})
\cite{Xiao2016FOP}. In the weak spin-orbital coupling limit or large Fermi
energy limit, one finds that $\tau_{\eta}^{L}\left(  \epsilon_{F}\right)
/\tau-1\simeq-\alpha k_{F}/2\epsilon_{F}$ while $\tilde{\tau}_{\eta}%
^{L}\left(  \epsilon_{F}\right)  /\tau-1\simeq-\alpha k_{F}/4\epsilon_{F}$,
where $k_{F}=\sqrt{2m\epsilon_{F}}/\hbar$. Figure 2(e) and 2(f) show that, in
the nonmagnetic Rashba 2DEG, Eq. (\ref{tau-p-multi}) (some times called the
\textquotedblleft modified relaxation time approximation\textquotedblright%
\ \cite{Trushin2009}) may be regarded as a more precise approximation than the
constant relaxation time approximation $1/\tau=\sum_{\eta^{\prime}}\int
\frac{d\phi^{\prime}}{2\pi}\omega_{\eta\phi,\eta^{\prime}\phi^{\prime}}^{2s}$
in the presence of short-range disorder potential.

It has been previously realized that the hybrid skew scattering contribution
(\ref{AHC}) obtained by the multi-band Keldysh formalism can be understood
within the SBE approach \cite{Kovalev2008}. This was explained
semi-quantitatively in previous works by using the $1/\tau^{||}$%
\ $\&$\ $1/\tau^{\perp}$ solution: for $\epsilon_{F}>J_{ex}$, $\omega
_{\eta\phi,\eta^{\prime}\phi^{\prime}}^{a-3nG}\left(  \epsilon_{F}\right)  =0$
and $\omega_{\eta\phi,\eta^{\prime}\phi^{\prime}}^{a-4nG}\left(  \epsilon
_{F}\right)  \sim n_{im}V_{0}^{4}\sin\left(  \phi^{\prime}-\phi\right)
I\left(  \epsilon_{F}\right)  $, thus Eq. (\ref{comp-nG}) yields $\tilde{\tau
}_{\eta}^{sk-3nG}=0$ and $\tilde{\tau}_{\eta}^{sk-4nG}\sim I\left(
\epsilon_{F}\right)  /n_{im}$. On the other hand, as shown by Eq.
(\ref{comp-sk-1}) and figure 2(c) and 2(d), in the SBE formalism the accurate
result Eq. (\ref{AHC}) does not follow from the
$1/\tau^{||}$\ $\&$\ $1/\tau^{\perp}$ solution.

Finally we mention that, it has been proposed that the analytical solution to SBE (\ref{SBE-n-origin}) under the first Born approximation in general (isotropic and anisotropic) 2DEGs should be obtained following an integral equation approach instead of the $1/\tau^{||}$%
\ $\&$\ $1/\tau^{\perp}$ solution or other relaxation time approximation schemes \cite{Vyborny2009}. It can be verified that in isotropic 2DEGs our solution to SBE (\ref{SBE-n-origin}) (or Eq. (\ref{SBE-1-3})) in the first Born approximation is consistent with the integral equation approach.

In summary, we have presented a systematic study on how to solve the SBE
describing AHE in isotropic 2DEG with static impurities. For
single-Fermi-circle 2DEG we proved that the widely-used $1/\tau^{||}$%
\ $\&$\ $1/\tau^{\perp}$ solution is correct for longitudinal and skew
scattering transport. While, for multiple-Fermi-circle 2DEG it was shown that
the solution of SBE could differ from the $1/\tau^{||}$\ $\&$\ $1/\tau^{\perp
}$ solution if a decoupling condition is unsatisfied. The decoupling
condition states in what cases the transport relaxation times (for
longitudinal transport and for skew scattering) on different bands are
decoupled to each other. We proved that this decoupling condition is
unsatisfied in a spin-polarized Rashba 2DEG with pointlike potential impurities
when both Rashba bands are partially occupied. Focusing on this specific
system, when considering intrinsic and hybrid skew scatterings, it was showed that the deviation of $1/\tau^{||}$\ $\&$\ $1/\tau^{\perp}$ solution from our solution is notable for large Fermi energies. For
coordinate-shift effects, the side-jump velocity acquired in the inter-band
elastic scattering process was shown to be more important for larger Rashba
spin-orbit coupling. The inter-band side-jump velocity may even exceed the
intra-band one for the outer Rashba band. The coordinate-shift contribution to
the AHE in the presence of pointlike impurities is significantly different
from that in the limit of smooth disorder potential analyzed by Sinitsyn et
al. \cite{Sinitsyn2005}. In the latter case the coordinate-shift contribution
has the same sign as the intrinsic one for sufficiently large Fermi energy,
while in the former case the coordinate-shift contribution has the opposite
sign as and is two times larger in magnitude than the intrinsic one. The
validity of our solution to the SBE is confirmed by the fact that the
calculated result for AHE is in full agreement with the multi-band Keldysh
formalism \cite{Kovalev2008,Kovalev2010}.

\begin{acknowledgments}
The authors greatly acknowledge the insightful discussions with Qian Niu, Karel Vyborny and Maxim Trushin. Especially we are grateful to Karel Vyborny for critical reading of the revised manuscript and providing useful suggestions. This Work is supported by National Natural Science Foundation of China (No. 11274013 and No. 11274018),
and NBRP of China (2012CB921300).
\end{acknowledgments}

\appendix
\begin{widetext}
\section{Calculations of the anti-symmetric part of scattering rates
responsible for skew scatterings}
\subsection{Truncating the T-Matrix expansion and Born approximations beyond
the 1st order}
The T-matrix expansion according to Lippmann-Schwinger equation is%
\begin{equation}
T_{l,l\prime}=V_{l,l\prime}+\sum_{l\prime\prime}\frac{V_{ll\prime\prime
}V_{l\prime\prime l\prime}}{\epsilon_{l}-\epsilon_{l\prime\prime}+i\epsilon
}+\sum_{l\prime\prime}\sum_{l\prime\prime\prime}\frac{V_{ll\prime\prime
}V_{l\prime\prime l\prime\prime\prime}V_{l\prime\prime\prime l\prime}}{\left(
\epsilon_{l}-\epsilon_{l\prime\prime}+i\epsilon\right)  \left(  \epsilon
_{l}-\epsilon_{l\prime\prime\prime}+i\epsilon\right)  }+...\text{\ \ \ }%
\end{equation}
For elastic scattering, the golden rule yields the scattering rate as
$\omega_{l,l\prime}=\omega_{l,l\prime}^{\left(  2\right)  }+\omega_{l,l\prime
}^{\left(  3\right)  }+\omega_{l,l\prime}^{\left(  4\right)  }+..$, where
$\omega_{l,l\prime}^{\left(  2\right)  }=\omega_{l,l\prime}^{2s}=\frac{2\pi
n_{im}V_{0}^{2}}{\hbar}\left\vert \langle u_{l\prime}|u_{l}\rangle\right\vert
^{2}\delta\left(  \epsilon_{l}-\epsilon_{l\prime}\right)  $ is symmetric,
$\omega_{l,l\prime}^{\left(  3\right)  }$ is related to non-Gaussian disorder
correlation \cite{Sinitsyn2007,Borunda2007}
\begin{equation}
\omega_{l,l\prime}^{\left(  3\right)  }=\frac{2\pi}{\hbar}\left(
\sum_{l\prime\prime}\frac{\left\langle V_{l\prime l}V_{ll\prime\prime
}V_{l\prime\prime l\prime}\right\rangle _{c}}{\epsilon_{l}-\epsilon
_{l\prime\prime}+i\epsilon}+c.c\right)  \delta\left(  \epsilon_{l\prime
}-\epsilon_{l}\right)  ,
\end{equation}
whereas $\omega_{l,l\prime}^{\left(  4\right)  }$ contains both Gaussian
$\left(  G\right)  $ and non-Gaussian $\left(  nG\right)  $ contributions
$\omega_{l,l\prime}^{\left(  4\right)  }=\omega_{l,l\prime}^{4G}%
+\omega_{l,l\prime}^{4nG}$. Here we only take into account the Gaussian part
responsible for the intrinsic skew scattering in the non-crossing
approximation \cite{Sinitsyn2007,Kovalev2010,Yang2011} and the non-Gaussian
part responsible for the hybrid skew scattering containing only one scattering
center \cite{Kovalev2008,Kovalev2009}. Then $\omega_{l,l\prime}^{4G}%
=\omega_{l,l\prime}^{4G,1}+\omega_{l,l\prime}^{4G,2}+\omega_{l,l\prime}%
^{4G,3}$ with
\begin{equation}
\omega_{l,l\prime}^{4G,1}=\frac{2\pi}{\hbar}\left(  \sum_{l\prime\prime}%
\sum_{l\prime\prime\prime}\frac{\left\langle V_{l\prime\prime\prime
l}V_{ll\prime\prime}\right\rangle _{c}}{\epsilon_{l}-\epsilon_{l\prime\prime
}+i\epsilon}\frac{\left\langle V_{l\prime\prime l\prime}V_{l\prime
l\prime\prime\prime}\right\rangle _{c}}{\epsilon_{l}-\epsilon_{l\prime
\prime\prime}-i\epsilon}\right)  \delta\left(  \epsilon_{l\prime}-\epsilon
_{l}\right)  ,
\end{equation}%
\begin{equation}
\omega_{l,l\prime}^{4G,2}=\frac{2\pi}{\hbar}\left(  \sum_{l\prime\prime}%
\sum_{l\prime\prime\prime}\frac{\left\langle V_{l\prime\prime l}V_{ll\prime
}\right\rangle _{c}}{\epsilon_{l}-\epsilon_{l\prime\prime\prime}-i\epsilon
}\frac{\left\langle V_{l\prime l\prime\prime\prime}V_{l\prime\prime\prime
l\prime\prime}\right\rangle _{c}}{\epsilon_{l}-\epsilon_{l\prime\prime
}-i\epsilon}+c.c.\right)  \delta\left(  \epsilon_{l\prime}-\epsilon
_{l}\right)  ,
\end{equation}%
\begin{equation}
\omega_{l,l\prime}^{4G,3}=\frac{2\pi}{\hbar}\left(  \sum_{l\prime\prime}%
\sum_{l\prime\prime\prime}\frac{\left\langle V_{ll\prime}V_{l\prime
l\prime\prime\prime}\right\rangle _{c}}{\epsilon_{l}-\epsilon_{l\prime
\prime\prime}-i\epsilon}\frac{\left\langle V_{l\prime\prime\prime
l\prime\prime}V_{l\prime\prime l}\right\rangle _{c}}{\epsilon_{l}%
-\epsilon_{l\prime\prime}-i\epsilon}+c.c.\right)  \delta\left(  \epsilon
_{l\prime}-\epsilon_{l}\right)  ,
\end{equation}
and $\omega_{l,l\prime}^{4nG}=\omega_{l,l\prime}^{4nG,1}+\omega_{l,l\prime
}^{4nG,2}$ with
\begin{equation}
\omega_{l,l\prime}^{4nG,1}=\frac{2\pi}{\hbar}\sum_{l\prime\prime}\sum
_{l\prime\prime\prime}\frac{\left\langle V_{l\prime l\prime\prime\prime
}V_{l\prime\prime\prime l}V_{ll\prime\prime}V_{l\prime\prime l\prime
}\right\rangle _{c}}{\left(  \epsilon_{l}-\epsilon_{l\prime\prime}%
+i\epsilon\right)  \left(  \epsilon_{l}-\epsilon_{l\prime\prime\prime
}-i\epsilon\right)  }\delta\left(  \epsilon_{l\prime}-\epsilon_{l}\right)  ,
\label{4)}%
\end{equation}%
\begin{equation}
\omega_{l,l\prime}^{4nG,2}=\frac{2\pi}{\hbar}\left(  \sum_{l\prime\prime}%
\sum_{l\prime\prime\prime}\frac{\left\langle V_{l\prime l\prime\prime\prime
}V_{l\prime\prime\prime l\prime\prime}V_{l\prime\prime l}V_{ll\prime
}\right\rangle _{c}}{\left(  \epsilon_{l}-\epsilon_{l\prime\prime\prime
}-i\epsilon\right)  \left(  \epsilon_{l}-\epsilon_{l\prime\prime}%
-i\epsilon\right)  }+c.c.\right)  \delta\left(  \epsilon_{l\prime}%
-\epsilon_{l}\right)  . \label{5)}%
\end{equation}
The symmetric parts of $\omega_{l,l\prime}^{\left(  3\right)  }\ $and
$\omega_{l\prime,l}^{\left(  4\right)  }$ only renormalize the result of
$\omega_{l^{\prime},l}^{2s}$ and thus will not be considered further
\cite{Sinitsyn2007}, while the anti-symmetric parts lead to skew scattering.
Therefore the scattering rate is approximated as Eq.
(\ref{total scattering rate}).
\subsection{Conventional skew scattering}
The anti-symmetric part of 3rd-order scattering rate can be obtained by%
\begin{equation}
\omega_{l,l\prime}^{a-3nG}=\frac{4\pi n_{im}V_{1}^{3}}{\hbar}\delta\left(
\epsilon_{k}^{\eta}-\epsilon_{k^{\prime}}^{\eta^{\prime}}\right)  \sum
_{\eta^{\prime\prime}\mathbf{k}^{\prime\prime}}\mathrm{Im}\frac{1}%
{\epsilon_{k}^{\eta}-\epsilon_{k^{\prime\prime}}^{\eta^{\prime\prime}%
}+i\epsilon}\mathrm{Im}\left[  \langle u_{\mathbf{k}}^{\eta}|u_{\mathbf{k}%
^{\prime}}^{\eta^{\prime}}\rangle\langle u_{\mathbf{k}^{\prime}}^{\eta
^{\prime}}|u_{\mathbf{k}^{\prime\prime}}^{\eta^{\prime\prime}}\rangle\langle
u_{\mathbf{k}^{\prime\prime}}^{\eta^{\prime\prime}}|u_{\mathbf{k}}^{\eta
}\rangle\right]  , \label{scattering rate-3nG}%
\end{equation}
where straightforward calculation gives $\int\frac{d\phi^{\prime\prime}}{2\pi}\mathrm{Im}\langle u_{\mathbf{k}}^{\eta
}|u_{\mathbf{k}^{\prime}}^{\eta^{\prime}}\rangle\langle u_{\mathbf{k}^{\prime
}}^{\eta^{\prime}}|u_{\mathbf{k}^{\prime\prime}}^{\eta^{\prime\prime}}%
\rangle\langle u_{\mathbf{k}^{\prime\prime}}^{\eta^{\prime\prime}%
}|u_{\mathbf{k}}^{\eta}\rangle=-\frac{1}{4}\eta\eta^{\prime}\eta^{\prime
\prime}\sin\left(  \phi^{\prime}-\phi\right)  \sin\theta\sin\theta^{\prime
}\cos\theta^{\prime\prime} $.
Thus we finally get the energy-integrated 3rd order anti-symmetric scattering
rate in Eq. (\ref{scattering rate-3nG-1})
\subsection{Intrinsic skew scattering}
Firstly, the anti-symmetric part of $\omega_{l,l\prime}^{4G,1}$ is shown to
be
\begin{align*}
\omega_{l,l\prime}^{a-4G,1}  &  =-\frac{2\pi\left(  n_{im}V_{0}^{2}\right)
^{2}}{\hbar}\delta\left(  \epsilon_{k}^{\eta}-\epsilon_{k^{\prime}}%
^{\eta^{\prime}}\right)  \sum_{\eta^{\prime\prime}}\sum_{\eta^{\prime
\prime\prime}}\int\frac{k^{\prime\prime}dk^{\prime\prime}}{2\pi}%
\mathrm{Im}\frac{1}{\left(  \epsilon_{k}^{\eta}-\epsilon_{k^{\prime\prime}%
}^{\eta^{\prime\prime}}+i\epsilon\right)  \left(  \epsilon_{k}^{\eta}%
-\epsilon_{k^{\prime\prime}}^{\eta^{\prime\prime\prime}}-i\epsilon\right)  }\\
&  \times\mathrm{Im}\int\frac{d\phi^{\prime\prime}}{2\pi}\langle
u_{\mathbf{k}}^{\eta}|u_{\mathbf{k}^{\prime\prime}}^{\eta^{\prime\prime}%
}\rangle\langle u_{\mathbf{k}^{\prime\prime}}^{\eta^{\prime\prime}%
}|u_{\mathbf{k}^{\prime}}^{\eta^{\prime}}\rangle\langle u_{\mathbf{k}^{\prime
}}^{\eta^{\prime}}|u_{\mathbf{k}^{\prime\prime}}^{\eta^{\prime\prime\prime}%
}\rangle\langle u_{\mathbf{k}^{\prime\prime}}^{\eta^{\prime\prime\prime}%
}|u_{\mathbf{k}}^{\eta}\rangle .
\end{align*}
If $\eta^{\prime\prime\prime}=\eta^{\prime\prime}$, $\mathrm{Im}\left[
\langle u_{\mathbf{k}}^{\eta}|u_{\mathbf{k}^{\prime\prime}}^{\eta
^{\prime\prime}}\rangle\langle u_{\mathbf{k}^{\prime\prime}}^{\eta
^{\prime\prime}}|u_{\mathbf{k}^{\prime}}^{\eta^{\prime}}\rangle\langle
u_{\mathbf{k}^{\prime}}^{\eta^{\prime}}|u_{\mathbf{k}^{\prime\prime}}%
^{\eta^{\prime\prime\prime}}\rangle\langle u_{\mathbf{k}^{\prime\prime}}%
^{\eta^{\prime\prime\prime}}|u_{\mathbf{k}}^{\eta}\rangle\right]  =0$ and then
only the $\eta^{\prime\prime\prime}=-\eta^{\prime\prime}$
term contributes:
\begin{gather}
\omega_{l,l\prime}^{a-4G,1}=-\frac{2\pi\left(  n_{im}V_{0}^{2}\right)  ^{2}%
}{\hbar}\delta\left(  \epsilon_{k}^{\eta}-\epsilon_{k^{\prime}}^{\eta^{\prime
}}\right)  \sum_{\eta^{\prime\prime}\mathbf{k}^{\prime\prime}}\mathrm{Im}%
\frac{1}{\left(  \epsilon_{k}^{\eta}-\epsilon_{k^{\prime\prime}}^{\eta
^{\prime\prime}}+i\epsilon\right)  \left(  \epsilon_{k}^{\eta}-\epsilon
_{k^{\prime\prime}}^{-\eta^{\prime\prime}}-i\epsilon\right)  }\nonumber\\
\times\mathrm{Im}\left[  \langle u_{\mathbf{k}}^{\eta}|u_{\mathbf{k}%
^{\prime\prime}}^{\eta^{\prime\prime}}\rangle\langle u_{\mathbf{k}%
^{\prime\prime}}^{\eta^{\prime\prime}}|u_{\mathbf{k}^{\prime}}^{\eta^{\prime}%
}\rangle\langle u_{\mathbf{k}^{\prime}}^{\eta^{\prime}}|u_{\mathbf{k}%
^{\prime\prime}}^{-\eta^{\prime\prime}}\rangle\langle u_{\mathbf{k}%
^{\prime\prime}}^{-\eta^{\prime\prime}}|u_{\mathbf{k}}^{\eta}\rangle\right]  .
\label{4G-1}%
\end{gather}
Secondly, we find that it is convenient to introduce two quantities
$\omega_{l,l\prime}^{a-4G,2^{\prime}}$ and $\omega_{l,l\prime}^{a-4G,3^{\prime
}}$ by
\begin{equation}
\omega_{l,l\prime}^{a-4G,2\prime}=\frac{\omega_{l,l\prime}^{4G,2}%
-\omega_{l\prime,l}^{4G,3}}{2},\text{ }\omega_{l,l\prime}^{a-4G,3\prime}%
=\frac{\omega_{l,l\prime}^{4G,3}-\omega_{l\prime,l}^{4G,2}}{2}=-\omega
_{l\prime,l}^{a-4G,2\prime},%
\end{equation}
with%
\begin{gather}
\omega_{l,l\prime}^{a-4G,2\prime}=-\frac{4\pi\left(  n_{im}V_{0}^{2}\right)
^{2}}{\hbar}\delta\left(  \epsilon_{k}^{\eta}-\epsilon_{k^{\prime}}%
^{\eta^{\prime}}\right)  \sum_{\eta^{\prime\prime}\mathbf{k}^{\prime\prime}%
}\mathrm{Im}\frac{1}{\left(  \epsilon_{k^{\prime}}^{\eta^{\prime}}%
-\epsilon_{k^{\prime}}^{-\eta^{\prime}}-i\epsilon\right)  \left(  \epsilon
_{k}^{\eta}-\epsilon_{k^{\prime\prime}}^{\eta^{\prime\prime}}-i\epsilon
\right)  }\nonumber\\
\times\mathrm{Im}\left[  \langle u_{\mathbf{k}}^{\eta}|u_{\mathbf{k}^{\prime}%
}^{\eta^{\prime}}\rangle\langle u_{\mathbf{k}^{\prime}}^{\eta^{\prime}%
}|u_{\mathbf{k}^{\prime\prime}}^{\eta^{\prime\prime}}\rangle\langle
u_{\mathbf{k}^{\prime\prime}}^{\eta^{\prime\prime}}|u_{\mathbf{k}^{\prime}%
}^{-\eta^{\prime}}\rangle\langle u_{\mathbf{k}^{\prime}}^{-\eta^{\prime}%
}|u_{\mathbf{k}}^{\eta}\rangle\right]  . \label{4G-2'}%
\end{gather}
Straightforward calculations lead to
\begin{equation}
\mathrm{Im}\int\frac{d\phi^{\prime\prime}}{2\pi}\langle u_{\mathbf{k}%
}^{\eta}|u_{\mathbf{k}^{\prime\prime}}^{\eta}\rangle\langle u_{\mathbf{k}%
^{\prime\prime}}^{\eta}|u_{\mathbf{k}^{\prime}}^{\eta^{\prime}}\rangle\langle
u_{\mathbf{k}^{\prime}}^{\eta^{\prime}}|u_{\mathbf{k}^{\prime\prime}}^{-\eta
}\rangle\langle u_{\mathbf{k}^{\prime\prime}}^{-\eta}|u_{\mathbf{k}}^{\eta
}\rangle=\frac{1}{4}\eta^{\prime}\sin\left(  \phi^{\prime}-\phi\right)
\frac{\alpha k}{\Delta_{k}}\frac{\alpha k^{\prime}}{\Delta_{k^{\prime}}}%
\frac{J_{ex}}{\Delta_{k^{\prime\prime}}}, \label{parameter-1}%
\end{equation}%
\begin{equation}
\mathrm{Im}\int\frac{d\phi^{\prime\prime}}{2\pi}\langle u_{\mathbf{k}%
}^{\eta}|u_{\mathbf{k}^{\prime}}^{\eta^{\prime}}\rangle\langle u_{\mathbf{k}%
^{\prime}}^{\eta^{\prime}}|u_{\mathbf{k}^{\prime\prime}}^{\eta^{\prime\prime}%
}\rangle\langle u_{\mathbf{k}^{\prime\prime}}^{\eta^{\prime\prime}%
}|u_{\mathbf{k}^{\prime}}^{-\eta^{\prime}}\rangle\langle u_{\mathbf{k}%
^{\prime}}^{-\eta^{\prime}}|u_{\mathbf{k}}^{\eta}\rangle=-\frac{1}{4}\eta
\eta^{\prime}\sin\left(  \phi^{\prime}-\phi\right)  \frac{\alpha k}{\Delta
_{k}}\frac{\alpha k^{\prime}}{\Delta_{k^{\prime}}}\frac{\eta^{\prime\prime
}J_{ex}}{\Delta_{k^{\prime\prime}}}, \label{parameter-2}%
\end{equation}
then the energy-integrated scattering rate for the intrinsic skew scattering
read%
\begin{equation}
\omega_{\eta\phi,\eta^{\prime}\phi^{\prime}}^{a-4G,1}\left(  \epsilon\right)
=\frac{\hbar N_{\eta^{\prime}}\left(  \epsilon\right)  }{4N_{0}\tau^{2}}%
\eta\eta^{\prime}\sin\left(  \phi^{\prime}-\phi\right)  \frac{J_{ex}}%
{J_{ex}^{2}+2\epsilon_{R}\epsilon}\frac{\alpha^{2}k_{\eta}\left(
\epsilon\right)  k_{\eta^{\prime}}\left(  \epsilon\right)  }{\Delta_{\eta
}\left(  \epsilon\right)  \Delta_{\eta^{\prime}}\left(  \epsilon\right)  },
\end{equation}
and $\omega_{\eta\phi,\eta^{\prime}\phi^{\prime}}^{a-4G,2^{\prime}}\left(
\epsilon\right)  \sim\sin\left(  \phi^{\prime}-\phi\right)  \sum_{\eta
^{\prime\prime\prime}}\eta^{\prime\prime}\frac{N_{\eta^{\prime\prime}}\left(
\epsilon\right)  }{\Delta_{\eta^{\prime\prime}}\left(  \epsilon\right)  }=0$
as the vanishing of $\omega_{\eta\phi,\eta^{\prime}\phi^{\prime}}^{a-3nG}$.
Thus we also have $\omega_{\eta\phi,\eta^{\prime}\phi^{\prime}}%
^{a-4G,3^{\prime}}\left(  \epsilon\right)  =0$, and then the total
anti-symmetric scattering rate arising from the 4th-order Gaussian disorder
correlation is completely attributed to $\omega_{\eta\phi,\eta^{\prime}%
\phi^{\prime}}^{a-4G,1}$: $\omega_{\eta\phi,\eta^{\prime}\phi^{\prime}}%
^{a-4G}\left(  \epsilon\right)  =\omega_{\eta\phi,\eta^{\prime}\phi^{\prime}%
}^{a-4G,1}\left(  \epsilon\right)  $.
\subsection{Hybrid skew scattering}
The anti-symmetric part of the non-Gaussian scattering rate Eq. (\ref{4)})
reads%
\begin{align}
\omega_{l,l\prime}^{a-4nG,1} &  =-\frac{2\pi n_{im}V_{0}^{4}}{\hbar}%
\delta\left(  \epsilon_{k}^{\eta}-\epsilon_{k^{\prime}}^{\eta^{\prime}%
}\right)  \sum_{\eta^{\prime\prime}\mathbf{k}^{\prime\prime}}\sum
_{\eta^{\prime\prime\prime}\mathbf{k}^{\prime\prime\prime}}\mathrm{Im}\frac
{1}{\left(  \epsilon_{k}^{\eta}-\epsilon_{k^{\prime\prime\prime}}%
^{\eta^{\prime\prime\prime}}+i\epsilon\right)  \left(  \epsilon_{k}^{\eta
}-\epsilon_{k^{\prime\prime}}^{\eta^{\prime\prime}}-i\epsilon\right)
}\nonumber\\
&  \times\mathrm{Im}\left[  \langle u_{\mathbf{k}}^{\eta}|u_{\mathbf{k}%
^{\prime\prime\prime}}^{\eta^{\prime\prime\prime}}\rangle\langle
u_{\mathbf{k}^{\prime\prime\prime}}^{\eta^{\prime\prime\prime}}|u_{\mathbf{k}%
^{\prime}}^{\eta^{\prime}}\rangle\langle u_{\mathbf{k}^{\prime}}^{\eta
^{\prime}}|u_{\mathbf{k}^{\prime\prime}}^{\eta^{\prime\prime}}\rangle\langle
u_{\mathbf{k}^{\prime\prime}}^{\eta^{\prime\prime}}|u_{\mathbf{k}}^{\eta
}\rangle\right]  .
\end{align}
Straightforward evaluation gives
\begin{equation}
\mathrm{Im}\int\frac{d\phi^{\prime\prime}}{2\pi}\int\frac{d\phi^{\prime
\prime\prime}}{2\pi}\langle u_{\mathbf{k}}^{\eta}|u_{\mathbf{k}^{\prime
\prime\prime}}^{\eta^{\prime\prime\prime}}\rangle\langle u_{\mathbf{k}%
^{\prime\prime\prime}}^{\eta^{\prime\prime\prime}}|u_{\mathbf{k}^{\prime}%
}^{\eta^{\prime}}\rangle\langle u_{\mathbf{k}^{\prime}}^{\eta^{\prime}%
}|u_{\mathbf{k}^{\prime\prime}}^{\eta^{\prime\prime}}\rangle\langle
u_{\mathbf{k}^{\prime\prime}}^{\eta^{\prime\prime}}|u_{\mathbf{k}}^{\eta
}\rangle=\frac{1}{8}\eta\eta^{\prime}\sin\left(  \phi^{\prime}-\phi\right)
\sin\theta\sin\theta^{\prime}\left(  \eta^{\prime\prime\prime}\cos
\theta^{\prime\prime\prime}-\eta^{\prime\prime}\cos\theta^{\prime\prime
}\right)  ,\nonumber
\end{equation}
after some manipulations we get the energy-integrated scattering rate as%
\begin{align}
&  \omega_{\eta\phi,\eta^{\prime}\phi^{\prime}}^{a-4nG,1}\left(
\epsilon\right)  =\frac{2\pi n_{im}V_{0}^{4}N_{0}^{2}N_{\eta^{\prime}}\left(
\epsilon\right)  }{\hbar}\frac{1}{4}\eta\eta^{\prime}\sin\left(  \phi^{\prime
}-\phi\right)  \frac{\alpha^{2}k_{\eta}\left(  \epsilon\right)  k_{\eta
^{\prime}}\left(  \epsilon\right)  }{\Delta_{\eta}\left(  \epsilon\right)
\Delta_{\eta^{\prime}}\left(  \epsilon\right)  }\nonumber\\
&  \times\mathrm{Im}\left[  \sum_{\eta^{\prime\prime}}\eta^{\prime\prime}%
\int_{\eta^{\prime\prime}}d\epsilon^{\prime\prime}\frac{N_{\eta^{\prime\prime
}}\left(  \epsilon^{\prime\prime}\right)  }{N_{0}}\frac{J_{ex}}{\Delta
_{\eta^{\prime\prime}}\left(  \epsilon^{\prime\prime}\right)  \left(
\epsilon^{\prime\prime}-\epsilon+i\epsilon\right)  }\sum_{\eta^{\prime
\prime\prime}}\int_{\eta^{\prime\prime\prime}}d\epsilon^{\prime\prime\prime
}\frac{N_{\eta^{\prime\prime\prime}}\left(  \epsilon^{\prime\prime\prime
}\right)  }{N_{0}}\frac{1}{\left(  \epsilon^{\prime\prime\prime}%
-\epsilon-i\epsilon\right)  }\right]  .
\end{align}
Note that the notation $\int_{\eta^{\prime\prime}}$ is to emphasize the
different lower integral limits of energy integral for different Rashba bands.
Here we assume $J_{ex}\geq\epsilon_{R}$, thus $\int_{+}=\int_{J_{ex}}^{\infty
}$, $\int_{-}=\int_{-J_{ex}}^{\infty}$ and then $\sum_{\eta^{\prime\prime}%
}\eta^{\prime\prime}\int_{\eta^{\prime\prime}}d\epsilon^{\prime\prime}%
=-\int_{-J_{ex}}^{J_{ex}}d\epsilon^{\prime\prime}$. Then%
\begin{gather*}
\omega_{\eta\phi,\eta^{\prime}\phi^{\prime}}^{a-4nG,1}\left(  \epsilon\right)
=\frac{2\pi n_{im}^{2}V_{0}^{4}N_{0}^{2}N_{\eta^{\prime}}\left(
\epsilon\right)  }{\hbar n_{im}}\frac{\eta\eta^{\prime}}{4}\sin\left(
\phi^{\prime}-\phi\right)  \frac{\alpha^{2}k_{\eta}\left(  \epsilon\right)
k_{\eta^{\prime}}\left(  \epsilon\right)  }{\Delta_{\eta}\left(
\epsilon\right)  \Delta_{\eta^{\prime}}\left(  \epsilon\right)  }%
\mathrm{Im}\left\{  -\int_{-J_{ex}}^{J_{ex}}d\epsilon^{\prime\prime}%
\frac{J_{ex}}{\sqrt{\epsilon_{R}^{2}+J_{ex}^{2}+2\epsilon_{R}\epsilon
^{\prime\prime}}\left(  \epsilon^{\prime\prime}-\epsilon+i\epsilon\right)
}\right.  \\
\times\left.  \left[  \int_{-J_{ex}}^{J_{ex}}d\epsilon^{\prime\prime\prime
}\frac{N_{-}\left(  \epsilon^{\prime\prime\prime}\right)  }{N_{0}}\frac
{1}{\left(  \epsilon^{\prime\prime\prime}-\epsilon-i\epsilon\right)  }%
+\int_{J_{ex}}^{\infty}d\epsilon^{\prime\prime\prime}\frac{2}{\left(
\epsilon^{\prime\prime\prime}-\epsilon-i\epsilon\right)  }\right]  \right\}  ,
\end{gather*}
by using $J_{ex}\geq\epsilon_{R}$ and $\epsilon>J_{ex}$, we get Eq.
(\ref{scattering rate 4nG}) with a minus sign where $I\left(  \epsilon\right)
$ represents%
\begin{equation}
I\left(  \epsilon\right) =\int_{-J_{ex}}^{J_{ex}}d\epsilon^{\prime\prime
}\frac{J_{ex}}{\sqrt{\epsilon_{R}^{2}+J_{ex}^{2}+2\epsilon_{R}\epsilon
^{\prime\prime}}\left(  \epsilon^{\prime\prime}-\epsilon\right)  }
=\frac{m}{\hbar^{2}}\frac{4J_{ex}%
}{k_{-}^{2}\left(  \epsilon\right)  -k_{+}^{2}\left(  \epsilon\right)  }%
\ln\frac{k_{+}^{2}\left(  \epsilon\right)  }{k_{-}^{2}\left(  \epsilon\right)
}.\label{I}
\end{equation}
We note that it is the difference in the lower integral limit of the energy
integral between the $\pm$ Rashba bands that contributes this $I\left(
\epsilon\right)  $, which is a key factor of hybrid skew scattering. The
calculation of the anti-symmetric scattering rate of the non-Gaussian
scattering rate Eq. (\ref{5)}) is similar, the factor $I\left(  \epsilon
\right)  $ also plays a key role. The resulting energy-integrated scattering
rate is $\omega_{\eta\phi,\eta^{\prime}\phi^{\prime}}^{a-4nG,2}\left(
\epsilon\right)  =-2\omega_{\eta\phi,\eta^{\prime}\phi^{\prime}}%
^{a-4nG,1}\left(  \epsilon\right)  $, then $\omega_{\eta\phi,\eta^{\prime}%
\phi^{\prime}}^{a-4nG}=\omega_{\eta\phi,\eta^{\prime}\phi^{\prime}}%
^{a-4nG,1}+\omega_{\eta\phi,\eta^{\prime}\phi^{\prime}}^{a-4nG,2}%
=-\omega_{\eta\phi,\eta^{\prime}\phi^{\prime}}^{a-4nG,1}$.
\end{widetext}


\bigskip

\begin{thebibliography}{99}                                                                                               %


\bibitem {Nagaosa2010}N. Nagaosa, J. Sinova, S. Onoda, A. H. MacDonald, and N.
P. Ong, Rev. Mod. Phys. \textbf{82}, 1539 (2010).

\bibitem {Niu2010}D. Xiao, M. C. Chang, and Q. Niu, Rev. Mod. Phys.
\textbf{82}, 1959 (2010).

\bibitem {Sinitsyn2007}N. A. Sinitsyn, A. H. MacDonald, T. Jungwirth, V. K.
Dugaev, and J. Sinova, Phys. Rev. B \textbf{75}, 045315 (2007).

\bibitem {Kovalev2010}A. A. Kovalev, J. Sinova, and Y. Tserkovnyak, Phys. Rev.
Lett. \textbf{105}, 036601 (2010).

\bibitem {SinitsynPRL2006}N. A. Sinitsyn, J. E. Hill, H. Min, J. Sinova, and
A. H. MacDonald, Phys. Rev. Lett. \textbf{97}, 106804 (2006).

\bibitem {Yang2011}S. A. Yang, H. Pan, Y. Yao, and Q. Niu, Phys. Rev. B
\textbf{83}, 125122 (2011).

\bibitem {Culcer2003}D. Culcer, A. H. MacDonald, and Q. Niu, Phys. Rev. B
\textbf{68}, 045327 (2003).

\bibitem {Inoue2006}J. I. Inoue, T. Kato, Y. Ishikawa, H. Itoh, G. E. W.
Bauer, and L. W. Molenkamp, Phys. Rev. Lett. \textbf{97}, 046604 (2006).

\bibitem {Dugaev2005}V. K. Dugaev, P. Bruno, M. Taillefumier, B. Canals, and
C. Lacroix, Phys. Rev. B \textbf{71}, 224423 (2005).

\bibitem {Sinitsyn2005}N. A. Sinitsyn, Q. Niu, J. Sinova, and K. Nomura, Phys.
Rev. B \textbf{72}, 045346 (2005).

\bibitem {Liu}S. Y. Liu and X. L. Lei, Phys. Rev. B \textbf{72}, 195329
(2005); S. Y. Liu, N. J. M. Horing, and X. L. Lei, Phys. Rev. B \textbf{74},
165316 (2006).

\bibitem {Onoda}S. Onoda, N. Sugimoto, and N. Nagaosa, Phys. Rev. Lett.
\textbf{97}, 126602 (2006); S. Onoda, N. Sugimoto, and N. Nagaosa, Phys. Rev.
B \textbf{77}, 165103 (2008).

\bibitem {Borunda2007}M. F. Borunda, T. S. Nunner, T. Luck, N. A. Sinitsyn, C.
Timm, J. Wunderlich, T. Jungwirth, A. H. MacDonald, and J. Sinova, Phys. Rev.
Lett. \textbf{99}, 066604 (2007).

\bibitem {Nunner2007}T. S. Nunner, N. A. Sinitsyn, M. F. Borunda, V. K.
Dugaev, A. A. Kovalev, Ar. Abanov, C. Timm, T. Jungwirth, J. I. Inoue, A. H.
MacDonald, and J Sinova Phys. Rev. B \textbf{76}, 235312 (2007).

\bibitem {Nunner2008}T. S. Nunner, G. Zarand, and F. vonOppen, Phys. Rev.
Lett. \textbf{100}, 236602 (2008).

\bibitem {Kovalev2008}A. A. Kovalev, K. Vyborny, and J. Sinova, Phys. Rev. B
\textbf{78}, 041305(R) (2008).

\bibitem {Kovalev2009}A. A. Kovalev, Y. Tserkovnyak, K. Vyborny, and J.
Sinova, Phys. Rev. B \textbf{79}, 195129 (2009).

\bibitem {Sinitsyn2008}N. A. Sinitsyn, J. Phys.: Condens. Matter \textbf{20},
023201 (2008).

\bibitem {Sinitsyn2006}N. A. Sinitsyn, Q. Niu, and A. H. MacDonald, Phys. Rev.
B 73, 075318 (2006).

\bibitem {Czaja2014}P. Czaja, F. Freimuth, J. Weischenberg, S. Blugel, and Y.
Mokrousov, Phys. Rev. B \textbf{89}, 014411 (2014).

\bibitem {Ebert2010}S. Lowitzer, D. Kodderitzsch, and H. Ebert, Phys. Rev.
Lett. \textbf{105}, 266604 (2010).

\bibitem {Ebert2015}D. Kodderitzsch, K. Chadova, and H. Ebert, Phys. Rev. B
\textbf{92}, 184415 (2015).

\bibitem {Hou2015}D. Hou, G. Su, Y. Tian, X. Jin, S. A. Yang, and Q. Niu,
Phys. Rev. Lett. \textbf{114}, 217203 (2015).

\bibitem {Schliemann2003}J. Schliemann and D. Loss, Phys. Rev. B \textbf{68},
165311 (2003).

\bibitem {Vyborny2009}The $1/\tau^{||}$\ $\&$\ $1/\tau^{\perp}$ solution was
initially proposed as an analytical solution for the SBE under the first Born
approximation in anisotropic 2DEG, in Ref. \onlinecite{Schliemann2003}.
However, subsequent researches have revealed that this solution does not
provide a reliable account of anisotropic transport in substantial cases. For a
detailed analysis, see, for example, K. Vyborny, A. A. Kovalev, J. Sinova, and
T. Jungwirth, Phys. Rev. B \textbf{79}, 045427 (2009). In appendix D of this reference, the authors also
stated that the $1/\tau^{||}$\ $\&$\ $1/\tau^{\perp}$ solution is valid in
isotropic 2DEG even beyond the first Born order. Here in this article we only
generally confirm this statement in isotropic single-Fermi-circle 2DEG, while
in 2DEG with multiple Fermi circles this statement may not work when the
decoupling conditions (\ref{exist-s}) and (\ref{exist-a}) are not satisfied.

\bibitem {Luttinger1958}J. M. Luttinger, Phys. Rev. \textbf{112}, 739 (1958).

\bibitem {Xiao2016FOP}C. Xiao, D. Li, and Z. Ma, Front. Phys., \textbf{11},
117201 (2016)

\bibitem {note}Eq. (15) in Ref. \onlinecite{Kovalev2008} has a typo: a factor
$m^{3}$ was missing in the numerator. $m$ is the effective in-plane mass of
Rashba electrons.

\bibitem {note-1}There are some typos in Eq. (12) of Ref.
\onlinecite{Borunda2007}. Here Eq. (\ref{comp-L}) avoids these typos.

\bibitem {Trushin2009}M. Trushin, K. Vyborny, P. Moraczewski, A. A. Kovalev,
J. Schliemann, and T. Jungwirth, Phys. Rev. B \textbf{80}, 134405 (2009).
\end{thebibliography}
\end{document}